%% file: main.tex
\documentclass[letterpaper,twocolumn,10pt]{article}
\usepackage{usenix-2020-09}

\usepackage{fancyvrb}

\usepackage{tikz}
\usepackage{amsmath}
\usepackage{amsthm}
\usepackage{listings}
\usepackage{booktabs}
\usepackage{mdframed}
\usepackage{pifont}
\usepackage{soul}
\usepackage{stfloats}
\usepackage{amssymb}
\usepackage{subcaption}
\usepackage{booktabs}
\usepackage{multirow}
\usepackage{xcolor}
\usepackage{colortbl}
\usepackage{fontawesome}
\usepackage{calc}
\usepackage[available,functional,reproduced]{usenixbadges}

\definecolor{codegreen}{rgb}{0,0.6,0}
\definecolor{codegray}{rgb}{0.5,0.5,0.5}
\definecolor{codepurple}{rgb}{0.58,0,0.82}
\definecolor{backcolour}{rgb}{0.95,0.95,0.92}
\lstdefinestyle{mystyle}{
    basicstyle=\fontsize{8}{8.4}\selectfont\ttfamily,
    commentstyle=\color{codegreen},
    stringstyle=\color{codepurple},
    keywordstyle=\color{red}, 
    numbers=left, %
    numbers=right, %
    numbers=none, %
    numberstyle=\tiny\color{codegray}, %
    stepnumber=2, %
    numbersep=5pt, %
    breakatwhitespace=false,         %
    breaklines=true,                 %
    breakindent=0pt,
    captionpos=b,                   %
    keepspaces=true,                   %
    showspaces=false,                
    showstringspaces=false, %
    showtabs=false,                  
    tabsize=2, %
    frame=single, %
    framerule=0.5pt, %
    framesep=5pt, %
    rulecolor=\color{black}, %
}

\newmdtheoremenv[
  linecolor=black,
  linewidth=0.5pt,
  innertopmargin=0.5pt,     %
  innerbottommargin=5pt,  %
  innerleftmargin=10pt,    %
  innerrightmargin=10pt,   %
]{observation}{Observation}

\newcommand{\bulletitem}[1]{%
	\noindent{\bf \textbullet}\hspace{0.07in}\makebox[\widthof{ #1}][l]{ #1}%
}

\newcommand\eat[1]{}

\usepackage{eso-pic}
\begin{document}
\AddToShipoutPictureBG*{\AtPageLowerLeft{\raisebox{20pt}{\makebox[\paperwidth]{\fontsize{7}{8}\selectfont Originally published in the 23rd USENIX Symposium on Networked Systems Design and Implementation (NSDI '26), pp. 209--235, 2026.}}}}

\date{}

\title{KeepON: Supporting Deterministic Traffic on Standard NICs}

\author{
{\rm Chuanyu Xue}$^*$ \quad {\rm Tianyu Zhang}$^\dagger$ \quad {\rm Andrew Loveless}$^\ddagger$ \quad {\rm Song Han}$^*$ \\[0.3em]
{$^*$University of Connecticut \quad $^\dagger$University of Iowa \quad $^\ddagger$NASA Johnson Space Center}
}

\maketitle

\vspace{-2em}  %
\begin{abstract}
Networked mission-critical applications ({\em e.g.}, avionics control and industrial automation) demand deterministic packet transmissions to meet stringent sensing and control timing requirements. While specialized infrastructures such as Time-Triggered Ethernet and Time-Sensitive Networking (TSN) ensure deterministic data delivery across switches, end devices still require specialized NICs ({\em e.g.}, TSN NICs or NVIDIA Mellanox) to eliminate endpoint indeterminism. However, deploying such NICs at every endpoint is costly and hinders compatibility with legacy systems.  To address this challenge, we propose KeepON, a novel software-based driver model that enables deterministic packet transmission on commodity NICs. The core idea is to continuously transmit fixed-size placeholder packets, establishing a predictable transmission pattern. Mission-critical packets are then precisely inserted into this stream by replacing placeholders at their scheduled transmission slots, ensuring timing accuracy. The placeholder packets are efficiently dropped at the first-hop switch, avoiding negative impacts on network performance.  We prototype KeepON by modifying the standard NIC driver of a Raspberry Pi\footnote{\url{https://github.com/ChuanyuXue/KeepON-rpi}}, and integrate it into a real-world TSN testbed. Experimental results show that KeepON achieves up to 130× improvement in scheduling accuracy compared to the default driver, and 2.1× improvement over a hardware-based solution.
\end{abstract}

\input{1-intro.tex}
\input{2-design.tex}
\input{3-driver.tex}

\input{4-sync.tex}

\input{5-modules.tex}
\input{6-evaluation.tex}
\input{7-case.tex}

\input{8-related.tex}

\input{ack}

\clearpage
\bibliographystyle{plain}
\bibliography{references}

\clearpage
\appendix
\input{9-appendix.tex}

\clearpage
\setcounter{secnumdepth}{0}

\input{10-ae-appendix.tex}

\clearpage
\end{document}

%% file: 1-intro.tex
\section{Introduction}
\label{sec:intro}

\begin{figure}[t]
    \centering
    \includegraphics[width=0.47\textwidth]{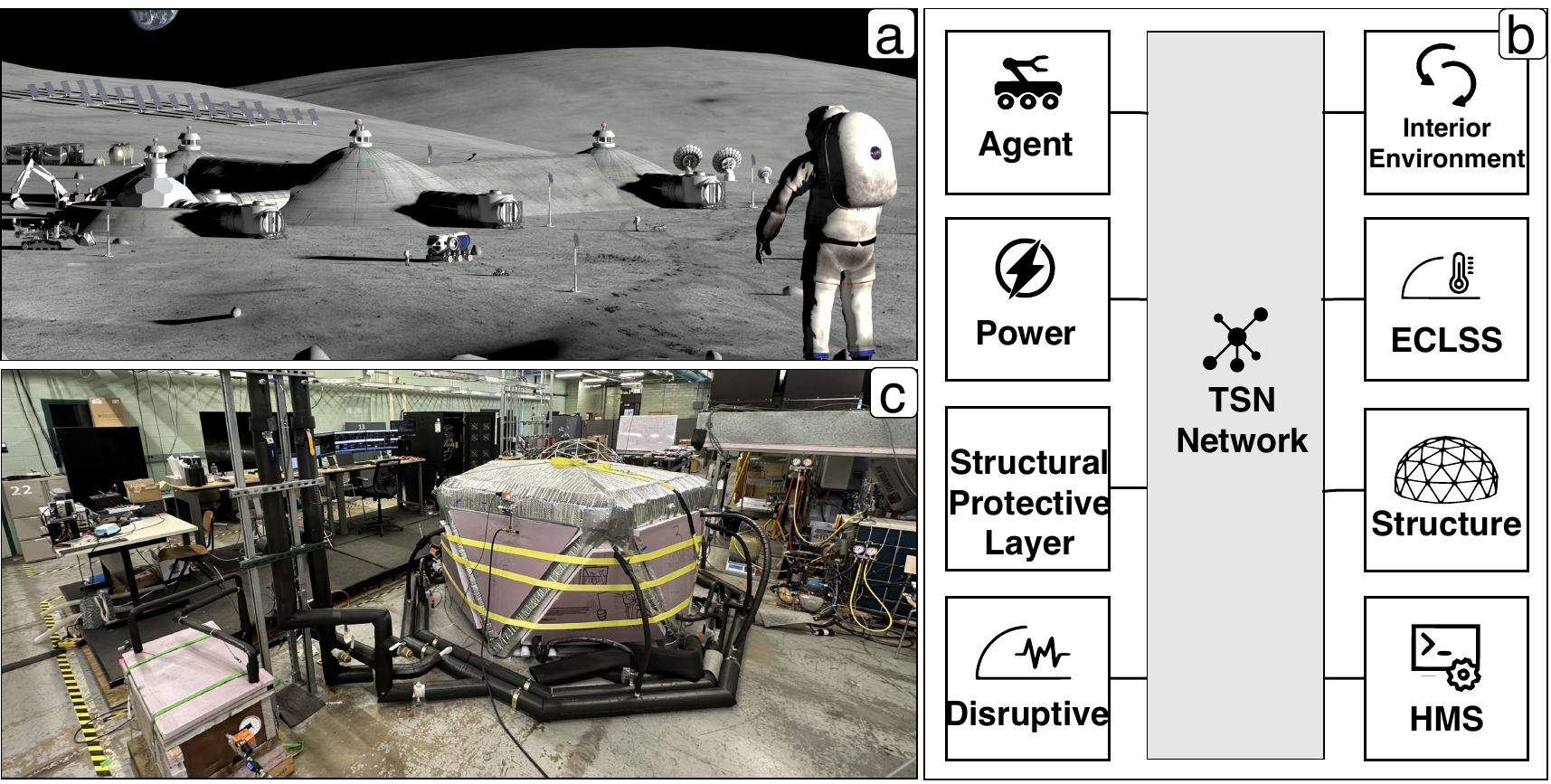}
    \caption{(a) Example of an extraterrestrial habitat system; (b) Subsystems of the habitat system interconnected by TSN; (c) Cyber-Physical Testbed with heterogeneous end devices~\cite{silva2025development}.}
    \vspace{-0.2in}
    \label{fig:motivation}
\end{figure}

Many networked mission-critical applications ({\em e.g.,} aircraft and spacecraft control~\cite{ieeedp, donnelly2022tsn, loveless2020ttegateway}, automotive systems~\cite{dai2020fixed, li2024data}, industrial manufacturing plants~\cite{ieee60802, zhang2024time}, and power generation facilities~\cite{tttech2022vestas,seewald2018iec61850}) impose stringent timing constraints on their sensing and control flows to ensure safe and reliable operations.
In these applications, correctness hinges not only on the functional delivery of data but also on strict timing guarantees, requiring packets to be transmitted within bounded delays and with minimal jitter. 
To meet such guarantees, various deterministic networking technologies have been developed ({\em e.g.,} Time-Sensitive Networking (TSN)~\cite{ieee8021q2018} , Time-Triggered Ethernet (TTE)~\cite{loveless2020ttegateway}, EtherCAT~\cite{IEC61158}, FlexRay~\cite{berwanger2001flexray}).  Among these, TSN has gained significant attention by integrating network-wide clock synchronization with time-aware scheduling and per-hop shaping, thereby enabling deterministic and reliable communication. However, in TSN systems, ensuring determinism solely at the network switches is insufficient; determinism at the end devices, particularly at the originating endpoints where packets are generated, is equally critical~\cite{donnelly2022tsn,barzegaran2022real, coleman2019emerging, xue2024towards}. 
As an illustrative example, we consider the design of NASA’s extra-terrestrial habitat system to highlight its system-level determinism requirements and demonstrate how improper behavior at end devices can disrupt overall system determinism as shown in Figure~\ref{fig:motivation}.

\vspace{0.02in}
\noindent {\bf Deterministic use case.} 
NASA’s Resilient Extra-Terrestrial Habitats Institute (RETHi) is an ongoing initiative aimed at developing technologies for resilient and autonomous habitat systems to support deep-space exploration. As a prerequisite to field deployment, the team is constructing a high-fidelity, hardware-in-the-loop emulator to evaluate the habitat’s control logic under realistic conditions. To this end, the RETHi team has developed {\em Cyber-Physical Testbed (CPT)}, a testbed emulating key subsystems essential for habitat resilience~\cite{silva2025development}, comprising both physical components and virtual elements in computing model, as shown in Figure~\ref{fig:motivation}(c).
Each subsystem is equipped with sensors, actuators, and computational units, and exchanges information with other subsystems in real time under strict timing constraints. For instance, the Environmental Control and Life Support System (ECLSS) periodically reports external pressure, thermal conditions, and other metrics to the Health Management System (HMS) every 75\,$\mu$s during crewed operations, while the Interior Environment Monitoring (IE) subsystem must report air pressure to ECLSS every 30\,$\mu$s in the event of an air leakage emergency.

To provide such timing guarantees, CPT employs TSN as its communication backbone, interconnecting subsystems through multiple TSN switches. Based on the flow specifications across subsystems ({\em e.g.}, period, deadline, source, and destination), the TSN network manager computes per-flow routes along with a network-wide schedule, realized as gate control lists (GCLs) on each switch. This schedule precisely dictates when each flow is released from its originating end device and how it traverses intermediate switches to ensure that end-to-end deadlines are satisfied.

While TSN switches can enforce schedules through hardware-based time gating, the habitat’s subsystems are deployed on heterogeneous physical and computing platforms tailored to their functionality, and some general-purpose hosts lack native timing support. For example, the CPT deployed at UConn integrates legacy devices (e.g., power supplies and sensors) through three Raspberry Pi 4B gateways (see the details in~\S\ref{sec:case-study}). Such timing misalignment at end devices introduces two critical issues: (1) frames may be released later than their GCL-mandated transmission times, leading to end-to-end deadline violations; and (2) frames released either earlier or later than expected can disrupt queue ordering at downstream switches, causing gate window violations that may cascade into a network-wide schedule failure~\cite{barzegaran2022real}. Consequently, TSN system determinism relies not only on deterministic forwarding within the switches but also on {\em transmission (TX) determinism} at the end devices~\cite{donnelly2022tsn, barzegaran2022real, coleman2019emerging, kyriakakis2020time, xue2024towards}.

\vspace{0.02in}
\noindent\textbf{Background.} To achieve TX determinism at end devices, existing approaches primarily rely on proprietary hardware support (e.g., Intel's LaunchTime feature on i210/i225 controllers~\cite{intel-i210-datasheet, IntelI225}, NVIDIA Mellanox's Accurate Scheduling technology~\cite{nvidia}), which offloads precise packet timing control from the operating system (OS) to the network interface card (NIC). This offloading circumvents the non-deterministic delays inherent at both the OS level ({\em e.g.}, dynamic socket buffer tuning, interrupts, and context switches~\cite{madden2019challenges, soares2010flexsc, xue2024towards}) and the architecture level ({\em e.g.}, PCIe contention, I/O operations, and batching mechanisms~\cite{neugebauer2018understanding, coleman2019emerging, xue2024towards, schuh2024cc, radhakrishnan2014senic, emmerich2018user}). Implementing this approach requires the NIC to provide two key capabilities: (1) an onboard hardware clock capable of high-precision synchronization via the IEEE 1588 Precision Time Protocol (PTP)~\cite{ieee1588}, and (2) dedicated logic to schedule packet transmissions based on timestamps derived from this synchronized clock~\cite{ieee60802, iic, ieeedp}. 

However, this hardware-dependent paradigm presents several significant limitations. First, the cost of deploying specialized NICs is prohibitively high, particularly in large-scale systems. For example, a modern spacecraft may include 50–100 end devices~\cite{amari2018specification, loveless2022impact}, and with per-NIC costs in the tens of thousands of dollars~\cite{tttech_aerospace_products}, the total expense for NICs alone can exceed \$1 million per system. Second, many industrial systems (e.g., civil avionics~\cite{mccormick2015certification}, smart grid~\cite{kumar2023review}, 	automotive~\cite{ISO26262:2018}) incorporate numerous legacy devices that are already certified, precluding the addition of new specialized NICs. Additionally, these systems often rely on small embedded devices without peripheral slots or expansion capabilities for hardware upgrades. Third, NICs with scheduling capabilities are scarce: only 13 of 341 surveyed Ethernet drivers (3.8\%~\S\ref{ssec:background_appendix}) support this feature, resulting in vendor lock-in and limited system flexibility.

\vspace{0.02in}
\noindent {\bf KeepON.} 
To address this gap, we propose KeepON, a software-based driver model that enables deterministic packet transmission on standard NICs without requiring specialized hardware. KeepON's key insight is that by maintaining continuous line-rate transmission of fixed-size dummy packets, we can create a stable and predictable timing reference directly in software. When data packets need transmission at specific times, KeepON replaces dummy packets at precise positions in the continuous stream, enabling deterministic transmission without hardware support. Dummy packets will be directly dropped by the first-hop switch via CRC check.  

Building on this insight, KeepON addresses three fundamental challenges. First, to enable precise and efficient packet insertion, we develop a replacement-based insertion mechanism alongside DMA-path optimization to minimize padding overhead. 
Second, to align end device local clocks with the absolute timings prescribed by the TSN network-wide schedule, we propose a PTP-based synchronization approach, further enhanced by our Dual Clock Mechanism to achieve sub-microsecond precision. 
Third, to support heterogeneous industrial traffic types, we design ring buffer partitioning that isolates real-time flows from best-effort traffic, ensuring that critical transmissions remain unaffected while residual capacity is used efficiently. 

\vspace{0.02in}
\noindent {\bf Results.} We prototype KeepON on Raspberry Pi 4B to demonstrate low-cost and generality, and evaluate it in both controlled experiments and a real-world NASA habitat testbed. Microbenchmarks demonstrate that KeepON achieves 3.7\,ns inter-arrival jitter at the MAC layer, representing a 130$\times$ improvement over the baseline GENET driver's 480.1\,ns, and outperforms even hardware-offloaded approaches, which achieve 7.7\,ns. In the case study of the habitat testbed, KeepON successfully maintains deterministic communication for mission-critical control loops, while the standard driver causes cascading timing failures.

%% file: 2-design.tex
\section{System Design}
\label{sec:observation}

\subsection{Dummy Packets-Enabled Precise Timing}
To enable KeepON to achieve TX determinism in a software-based manner without reliance on specialized hardware, a central requirement is a precise software-based timing mechanism at the end device. 
Existing literature~\cite{emmerich2015moongen, takano2005design, lee2013sonic} has explored the concept of using dummy packets to regulate packet transmission, where intentionally corrupted/unused packets are inserted into the transmission stream to space out data bursts or control inter-packet intervals. Inspired by this idea, we propose a novel technique that leverages dummy packets not only to shape the stream but to realize a time reference that allows the NIC to control the absolute transmission time and order of data packets so that they conform to the TSN schedule. 

Specifically, our approach involves instructing the NIC to continuously transmit fixed-size corrupted packets at the line rate, and thus, we can accurately measure elapsed time by counting the number of transmitted corrupted packets, effectively creating a software-based clock. When a data packet is scheduled for transmission, it is inserted into the transmission stream by replacing a corrupted packet at a specific position. In effect, the dummy stream acts as a metronome that advances in equal steps, and each data packet is inserted at the correct step so that its transmission time and relative order match the schedule. 

For this approach to be viable, two key conditions must be satisfied: 1) the NIC must sustain continuous transmission at the line rate with stable and predictable per-packet timing, and 2) corrupted packets must be discarded at the first hop to prevent them from flooding the network, and must not introduce additional overhead that degrades host/network performance. We validate these conditions with two experiments conducted on two machines directly connected via their Ethernet interfaces over a 1\,Gbps link, using different commodity NICs on the sender and an Intel i210 NIC on the receiver. 

\begin{figure}[tb]
    \centering
    \includegraphics[width=0.4\textwidth]{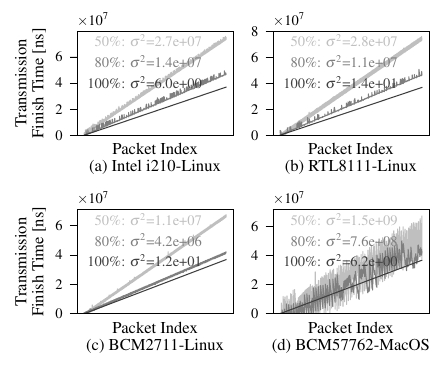}
    \vspace{-0.15in}
    \caption{Packet transmission finish time for fixed-size Ethernet frames under varying network loads across 4 different NIC/OS configurations.}\label{fig:insights}
    \vspace{-0.15in}
\end{figure}

In the first set of experiments, we employed a variety of common commodity NICs and OSs to measure the transmission finish times of fixed-size Ethernet frames (1500 bytes) under varying network utilization levels. As shown in Figure~\ref{fig:insights}, across all tested NIC and OS configurations, when the NIC operates below line rate ({\em e.g.}, at 50\% and 80\% utilization), the transmission finish times exhibit noticeable fluctuations. In contrast, when the NIC operates at 100\% line rate, the transmission finish times reveal a clear and consistent linear relationship with the packet index, indicating that the time intervals between successive transmissions become effectively constant and predictable. This behavior arises because at line rate, the NIC always has a frame ready for transmission, and both the per-frame transmission time and the inter-frame gap (IFG) are determined at the physical layer, eliminating any upper-level-induced timing variability.

\begin{figure}[tb]
    \centering
    \includegraphics[width=0.44\textwidth]{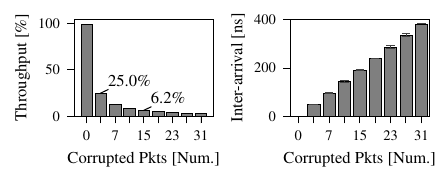}
    \vspace{-0.15in}
    \caption{Throughput and packets inter-arrival time at the receiver under varied numbers of corrupted packets.}\label{fig:insights2}
    \vspace{-0.15in}
\end{figure}

In the second set of experiments, we inserted varying numbers of corrupted packets into the data packet transmission stream and measured the RX throughput as well as the inter-arrival times of the received data packets. In Figure~\ref{fig:insights2} (left), the RX throughput matches exactly the fraction of data packets transmitted from the sender, indicating that all corrupted packets are effectively discarded. In Figure~\ref{fig:insights2} (right), the results demonstrate that, upon uniformly inserting different numbers of corrupted packets, the packet inter-arrival times at the receiver scale proportionally, confirming that the process of dropping corrupted packets incurs no additional timing overhead. 

\vspace{0.02in}
\noindent {\bf Distinctions from prior dummy-packet techniques.} 
Although prior work utilizes dummy packets for transmission regulation but falls short of achieving TX determinism, KeepON draws inspiration from it while introducing novelties in both the manner of dummy packet deployment and the capabilities achieved. 
First, prior work \textit{inserts} dummy packets of various sizes between data packets to enlarge inter-packet gaps and thereby space out bursts or enforce a target rate. In contrast, KeepON maintains a continuous stream of fixed-size dummy packets at the line rate and determines each data packet’s send time by \textit{replacing} the dummy packet at a specific position. Second, prior work does not provide precise control over the absolute packet send time nor allow arbitrary control of packet order. KeepON provides both: \textit{absolute} timing precision (through our proposed EPHC~\S\ref{ssec:EPHC} and synchronization mechanism~\S\ref{sec:sync}) and arbitrary packet ordering. 

\subsection{KeepON Architecture}

\begin{figure}[tb]
    \centering
    \includegraphics[width=0.48\textwidth]{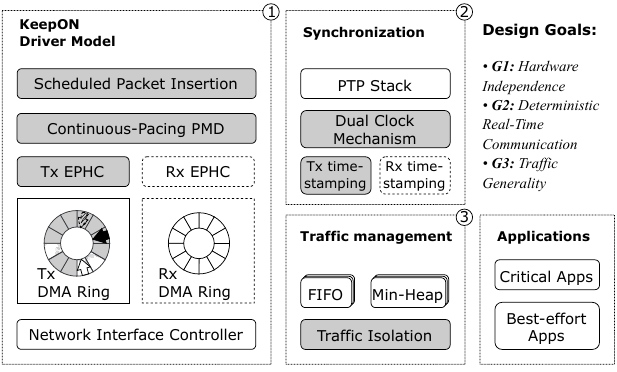}
    \caption{The overall architecture of KeepON with the grey blocks highlighting our newly designed components. Dashed-line blocks represent optional components for performance tuning ({\em e.g.}, RX-EPHC and RX timestamping).}\label{fig:TSN-arch}
    \vspace{-0.15in}
\end{figure}

Building on the dummy-packet–enabled precise timing mechanism, KeepON addresses several key challenges to realize software-based TX determinism, enabling its deployment in industrial applications. The overall architecture of KeepON is given in Figure~\ref{fig:TSN-arch}.

\vspace{0.02in}
\noindent \textit{Challenge 1: Hardware-independent precise local timing}. We design a KeepON driver model ({\ding{192}}) with three cooperating components: 1) the Continuous-Pacing Poll Mode Driver (CP-PMD) module, which maintains a continuous stream of transmissions at the line rate to establish a stable baseline for timing; 2) the Emulated PTP Hardware Clock (EPHC) module, which emulates a high-precision clock function by leveraging the predictable cadence of the dummy packet stream; and 3) the Scheduled Packet Insertion module, which precisely inserts data packets into the continuous stream by replacing designated dummy packets, thereby enforcing scheduled transmission times.

\vspace{0.02in}
\noindent \textit{Challenge 2: Aligning local time with the network schedule}. While the KeepON driver model establishes an accurate local clock at the end device, adhering to the absolute transmission times specified in the TSN network-wide schedule requires aligning this local clock with the global network time. To address this, we propose a Synchronization module (\ding{193}) that synchronizes the EPHC across multiple devices, utilizing lightweight protocols to achieve sub-microsecond accuracy and ensure coherent timing throughout the TSN system.

\vspace{0.02in}
\noindent \textit{Challenge 3: Supporting heterogeneous traffic}. Industrial applications carry both critical real-time flows with determinism requirements and non-critical best-effort flows~\cite{zhang2024survey, ieee60802, iic}. To support heterogeneous industrial traffic, we design a Traffic Management module (\ding{194}) that handles diverse traffic types through traffic isolation via Direct Memory Access (DMA) ring partitioning and traffic pre-buffering using dedicated queues for different flow types. 

In the following sections, we elaborate on the detailed design of each KeepON component.

%% file: 3-driver.tex
\begin{figure}[tb]
    \centering
    \includegraphics[width=0.52\textwidth]{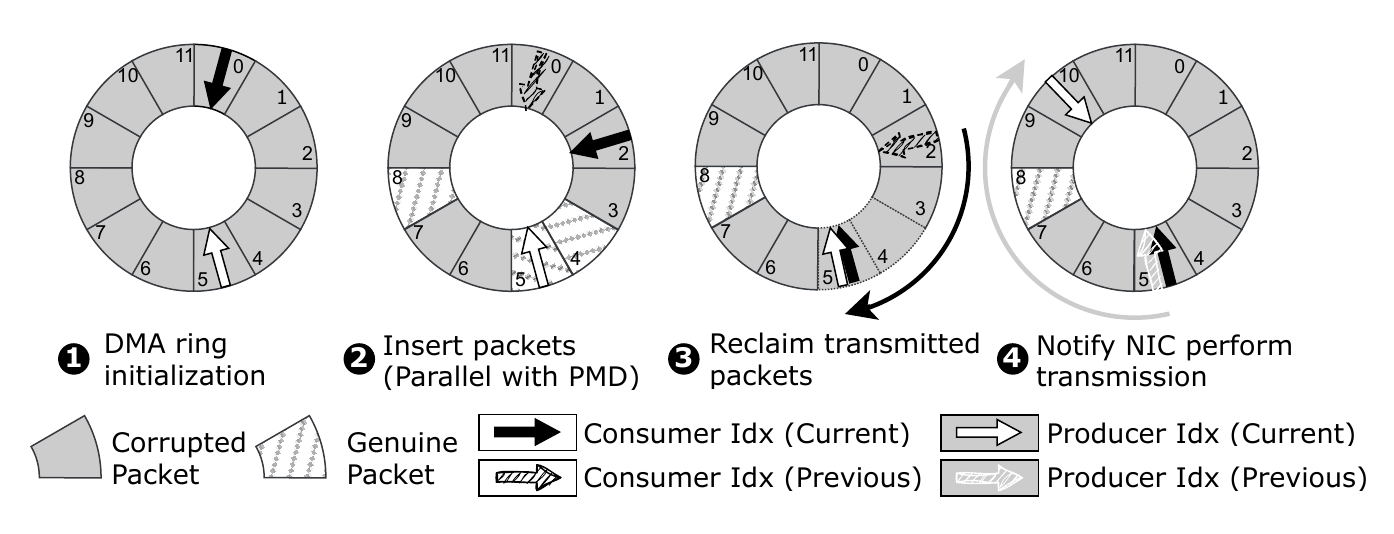}
    \vspace{-0.15in}
    \caption{DMA ring management operations in the KeepON driver model, illustrating both the Continuous-Pacing PMD (\S\ref{sec:continuous-pacing}) and Scheduled Packet Insertion (\S\ref{sec:scheduled-packet-insertion}).}\label{fig:ring}
    \vspace{-0.15in}
\end{figure}

\section{KeepON Driver Model}

\subsection{Continuous-Pacing PMD (CP-PMD)}
\label{sec:continuous-pacing}
Unlike conventional drivers that perform on-demand packet transmission, CP-PMD aims to enable constant NIC transmission by filling idle bandwidth with corrupted packets, operating over a pre-initialized ring buffer populated with descriptors for fixed-size corrupted frames.

Figure~\ref{fig:ring} shows an example of a DMA ring buffer with 12 descriptors (\ding{202}). It implements a persistent polling loop in three steps: \ul{1) Identify Transmitted Packets:} The driver queries the current consumer index, identifying descriptors whose transmission has been completed. 
For example, the consumer index (black arrow) moves from 2 to 5, signifying that 3, 4, and 5 are transmitted (\ding{204}); 
\ul{2): Reclaim and Reset Descriptors:} It resets the completed descriptors (4, 5) back to the placeholder state with corrupted-CRC (\ding{204}), marking them as available for subsequent overwrites by the packet insertion mechanism (to be discussed in \S\ref{sec:scheduled-packet-insertion}); 
\ul{3) Advance Producer Index:} PMD consistently advances the producer index by a certain batch size, informing the NIC how far it should continue to process descriptors in the ring buffer (\ding{205}). 
This continuous cycle of querying, resetting, and advancing indices ensures that the NIC hardware maintains continuous line-rate operation, regardless of whether any new application packets are inserted into the ring.

\begin{figure}[tb]
    \includegraphics[width=0.45\textwidth]{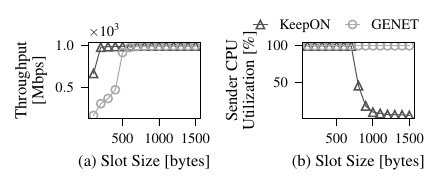}
    \vspace{-0.15in}
    \caption{Comparison of CPU utilization of KeepON and GENET under different batch size and slot size settings. The error bars represent standard deviations.}
    \label{fig:pmd_comparison} 
    \vspace{-0.15in}
\end{figure}

\vspace{0.02in}
\noindent \textbf{Parameter settings.} 
In CP-PMD, the setting of two parameters, {\em i.e.}, slot size and batch size, affects the system performance, {\em e.g.}, real-time performance, CPU usage, and clock accuracy. 
The slot size determines the time granularity of packet transmissions, where a smaller slot size enables finer-grained timing control but increases system overhead due to more frequent packet handling. 
The batch size specifies the number of descriptors that the producer index advances in each polling iteration ({\em e.g.}, batch size is 5 in Fig.~\ref{fig:ring} (\ding{205})). A larger batch size reduces per-packet processing costs, thus lowering CPU and PCIe overheads. But it comes at the expense of reduced time granularity, lower clock accuracy, and the flexibility in packet scheduling (immediate insertion window in \S\ref{sec:scheduled-packet-insertion}). 

\vspace{0.02in}
\noindent \textbf{Experimental validation.} We evaluate CP-PMD's continuous transmission efficiency by measuring receiver throughput and sender CPU utilization across slot sizes from 100 to 1500 bytes with a batch size fixed at 16. We compare CP-PMD (KeepON) against the conventional default driver (GENET), sending with raw socket with the same hardware settings. Figure~\ref{fig:pmd_comparison} shows that CP-PMD consistently outperforms the conventional approach. CP-PMD achieves line-rate ($\geq$ 99.9\% capacity) throughput at 300-byte packets and maintains CPU utilization $\leq$ 50\% for packets at 800 bytes. In contrast, GENET requires 700-byte packets to reach line-rate and exhibits full CPU utilization across all packet sizes due to its on-demand transmission model.

\subsection{Emulated PTP Hardware Clock}
\label{ssec:EPHC}
EPHC aims to provide a clock domain derived from the NIC's packet counter via mapping the desired time instant to a specific DMA ring slot. 

\vspace{0.02in}
\noindent \textbf{Clock specification.} EPHC maps the constant-rate physical layer activity, enforced by CP-PMD, to a time base. To realize the mapping, we employ a linear clock model comprising three key components. The current time is calculated as: cycle-count $\times$ cycle-period $+$ offset.

\vspace{0.02in}
\noindent $\bullet$ \textit{Cycle-count} is derived by the packet counter, which increments with each transmitted packet and serves as the "ticks" of the emulated clock.

\vspace{0.02in}
\noindent $\bullet$ \textit{Cycle-period} defines the duration of each transmission cycle and equals the division of slot size by line rate.

\vspace{0.02in}
\noindent $\bullet$ \textit{Offset} serves as the clock's epoch, aligning the origin of the emulated clock to a meaningful global time reference, such as UTC time~\cite{itu-t-g8271}. It is obtained in the initialization phase. 

\begin{figure}[tb]
    \includegraphics[width=0.45\textwidth]{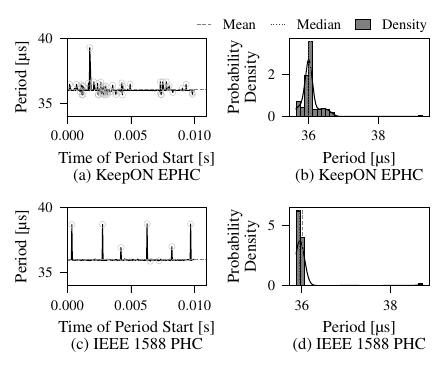}
    \vspace{-0.15in}
    \caption{Comparison of clock stability of KeepON's EPHC (Fig.(a)\&(b)) and IEEE 1588 hardware clock (Fig.(c)\&(d)), both targeting a 36\,$\mu$s period.} %
    \label{fig:gpio}
    \vspace{-0.15in}
\end{figure}

\vspace{0.02in}
\noindent \textbf{Clock stability measurements.} We compare the clock stability of the proposed EPHC with the Intel i210 NIC's hardware PTP clock (PHC). Since the software-emulated EPHC cannot be directly interfaced with measurement equipment, we assess stability by using each clock to toggle a GPIO pin, generating a square wave with a 36\,$\mu$s period (as a multiple of EPHC's clock granularity at 1500-byte slot size). The resulting periods are measured using a Tektronix TBS2000B oscilloscope.

As shown in Figure~\ref{fig:gpio}, the EPHC achieves a mean and median period closely matching the target, demonstrating comparable central tendency to the hardware clock. 
However, EPHC exhibits higher jitter, reflected in the wider distribution of its measured periods in Figure~\ref{fig:gpio}(b) compared to the tighter distribution for the PHC in Figure~\ref{fig:gpio}(d). This increased jitter primarily stems from the granularity limits of EPHC's packet-counting-based time estimation. Nevertheless, most of EPHC's jitter remains within $\leq$1\,$\mu$s of the mean, indicating relatively controlled variation.

\begin{figure}[tb]
    \includegraphics[width=0.45\textwidth]{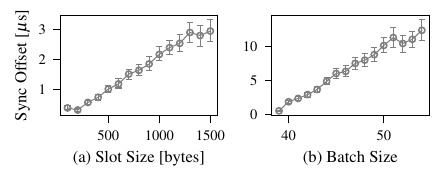}
    \vspace{-0.15in}
    \caption{Impact of slot size and batch size on the EPHC clock stability, evaluated by measuring the absolute time offset between the EPHC clock and system UTC clock. }\label{fig:offset}
    \vspace{-0.15in}
\end{figure}

We further investigate how the slot size and batch size settings affect the clock stability of EPHC by continuously measuring the absolute offset between the EPHC and the system UTC clock. 
Figure~\ref{fig:offset} shows that both a larger slot size and a larger batch size lead to larger clock offsets. The former is due to the larger clock granularity and the latter is caused by the less frequent clock update. 
In the experiments, we also notice that the offset is negligible ($\leq 1~\mu$s) when the batch size is set smaller than $39$.

\vspace{0.02in}
\noindent \textbf{Clock adjustment:} To enable the EPHC clock to synchronize with other clocks in the network, two types of clock adjustments are supported. 
1)~\textit{Offset Adjustment} aligns the EPHC’s time reference by either applying a fixed offset correction or directly setting the clock to a specified time. 
2)~\textit{Rate adjustment} modifies the cycle-period, effectively adjusting the emulated clock’s progression rate to correct for frequency drift and maintain long-term alignment with a reference clock.
These clock adjustments never modify the cycle count, which remains a monotonic counter tied solely to packet transmissions. This design principle ensures that transmission scheduling remains undisrupted during clock synchronization.

\begin{figure*}[tb]
    \centering
    \includegraphics[width=0.9\textwidth]{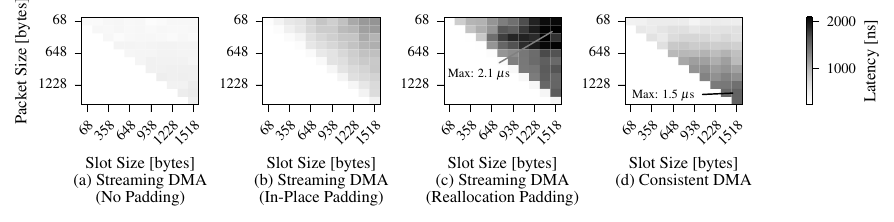}
    \vspace{-0.15in}
    \caption{Latency comparison for different padding strategies. Heatmaps show latency for padding (lighter is lower/better) based on packet size (y-axis) vs. target DMA slot size (x-axis).}\label{fig:dma-heatmap}
    \vspace{-0.15in}
\end{figure*}

\subsection{Scheduled Packet Insertion}
\label{sec:scheduled-packet-insertion}
Unlike traditional drivers, which process packet descriptors in an FIFO manner, KeepON performs scheduled packet insertion and overwrites placeholder descriptors at specific indices in the DMA ring to achieve scheduled transmissions. 

\vspace{0.02in}
\noindent \textbf{Insertion process.} 
When an outgoing packet arrives at the driver with its designated scheduled time ($t_{sch}$) in the EPHC clock domain, the driver calculates the target slot index as \mbox{(($t_{sch}$ $-$ offset) $/$ cycle-period) $\%$ $N$}, where $N$ is the DMA ring size, and performs three validation checks.
1) It verifies that the target slot is occupied by a placeholder descriptor associated with a corrupted CRC.
2) The target slot must be sufficiently ahead of the NIC's current consumer index by at least one batch size, and before the consumer index wraps around in the next cycle to maintain a safe temporal margin, {\em i.e.}, within immediate insertion window $[\text{c\_index} + \text{batch size},  \text{ c\_index} + N)$. This means a packet must arrive at the driver at least advance time of batch-size $\times$ slot-size before its scheduled transmission time.
3) It verifies that the incoming packet's application has ownership of the target slot (more details are discussed in \S\ref{ssec:isolation}).
If all the validation checks succeed, the driver pads the packet to match the slot size and inserts it into the ring buffer at the target slot index. For example, as shown in Figure~\ref{fig:ring}, packets are inserted at slots 4, 5, and 8  (\ding{203}), rather than being appended sequentially to the queue as in a conventional driver. 
After insertion, the driver updates the corresponding descriptor to mark the slot as valid by setting the CRC field to a correct value. 
If any validation check fails, the driver discards the packet and returns an error code to the application.

One challenge in packet insertion is padding outgoing packets to match fixed-size slots, which introduces significant memory write delays. 
Traditional drivers use zero-copy transmission with standard streaming DMA, directly mapping packet buffers (skbs) to the NIC's DMA engine and unmapping them after transmission. This incurs a small and size-independent per-packet map/unmap overhead ($\leq$ 0.5\,$\mu$s), as shown in Figure~\ref{fig:dma-heatmap}(a). 
However, zero-copy transmission loses its performance advantage when packets require padding, where the driver must write zeros to fill empty slot space, creating significant overhead for small packets in large slots, as shown in Figure~\ref{fig:dma-heatmap}(b). 
Even worse, if the skb lacks sufficient tailroom, a costly re-allocation, copy, padding (\texttt{memset}), and remapping sequence is triggered, adding significant and unpredictable latency. 
Figure~\ref{fig:dma-heatmap}(c) shows that this reallocation path often incurs latency exceeding 2\,$\mu$s. Tests on Linux kernel 6.12 with uniformly distributed packet sizes reveal how frequently this occurs: for UDP and raw sockets, 37.5\% of packets lack sufficient skb tailroom for slot padding, and for TCP, tailroom is always zero. 
This confirms that relying on streaming DMA for padding frequently triggers the costly reallocation behavior.

\vspace{0.02in}
\noindent \textbf{Packet padding.} 
To address the padding overhead, we propose to use consistent DMA by pre-allocating and permanently mapping fixed-size packet buffers that match the transmission slot size (see implementation details in~\S\ref{ssec:driver-impl}). When a new packet arrives, its payload is copied directly into the designated pre-mapped buffer, trading map/unmap overhead for a predictable memory copy cost. 
As shown in Figure~\ref{fig:dma-heatmap}(d), this approach keeps the packet insertion latency within $1.5~\mu$s, primarily determined by the payload size. 
More importantly, since the buffers are fixed-size and reused, padding (zeroing the unused portion) can be efficiently deferred to CP-PMD’s polling phase (\ding{204} in Figure~\ref{fig:ring}) when resetting descriptors. This shifts the memory write overhead off the critical packet insertion path, leaving only the payload \texttt{memcpy} cost during scheduling. 

\begin{figure}[tb]
    \centering
    \includegraphics[width=0.45\textwidth]{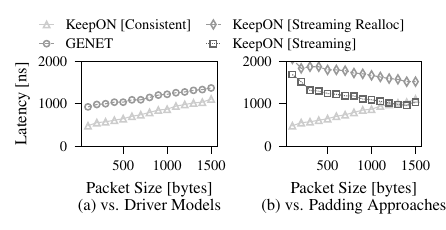}
    \vspace{-0.15in}
    \caption{Comparison of total driver transmit path latency. (a) Median delay of KeepON and GENET. (b) Median delay of different padding approaches, comparing streaming DMA (w/o sufficient tailroom) versus consistent DMA.}\label{fig:dma-latency}
    \vspace{-0.15in}
\end{figure}

We compare the total latency in the driver's transmit path, measured from the moment when a packet arrives at the driver to the time when it is handed over to the NIC. 
Figure~\ref{fig:dma-latency}(a) shows the median delay of the KeepON driver and GENET. KeepON consistently achieves lower latency, with an average delay of 7\,$\mu$s compared to 11\,$\mu$s for the baseline. This improvement primarily results from eliminating the DMA mapping/unmapping overhead and avoiding the IOMMU doorbell operation typically required to notify the NIC for transmission. By moving NIC notifications out of the critical data path (handled by the PMD), KeepON reduces transmission overhead and latency.
Figure~\ref{fig:dma-latency}(b) compares the impact of different padding strategies. When using streaming DMA (square-mark line), its performance is close to that of consistent DMA, when all incoming packets have sufficient tailroom for padding and large packets are handled ($\geq$ 1200 bytes). 
However, when tailroom is insufficient and reallocation is required, streaming DMA incurs significantly higher latency (diamond-mark line) due to padding, with an average delay of 17\,$\mu$s.

%% file: 4-sync.tex
\section{Network-Wide Synchronization}\label{sec:sync}
To achieve end-to-end deterministic packet transmission, the EPHC clock must synchronize across other nodes to enable network-wide synchronization.

\subsection{PTP Synchronization}
\begin{figure}[tb]
    \centering
    \includegraphics[width=0.42\textwidth]{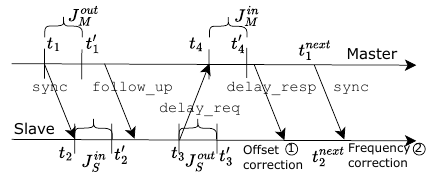}
    \vspace{-0.1in}
    \caption{Single-hop packet exchange and timestamping for PTP synchronization between master and slave demonstrating the calculation of clock offset and frequency adjustments.}
    \vspace{-0.15in}
    \label{fig:ptp}
\end{figure}

We synchronize the EPHC clock using the IEEE 1588 PTP~\cite{ieee1588}, which corrects both time offset and frequency drift relative to a master clock. 
As illustrated in Figure~\ref{fig:ptp}, PTP operates via a timed message exchange requiring four timestamps: \texttt{sync} sent by master at $t_1$, \texttt{sync} received by slave at $t_2$, \texttt{delay\_req} sent by slave at $t_3$, and \texttt{delay\_req} received by master at $t_4$. 
Assuming symmetric network delay~\cite{ieee1588}, the slave estimates the time offset each round of message exchange (\ding{192}) as $[(t_2 - t_1) - (t_4 - t_3)] / 2$, which is used to adjust the EPHC's \textit{Offset}. 
Frequency differences, estimated by observing the change in offset over time, {\em e.g.}, using two successive \texttt{sync} message timestamps $t_1/t_2$ and $t_1^{next}/t_2^{next}$, are used to adjust the EPHC's rate parameter (\ding{193}).

Obtaining precise timestamps ($t_1, t_2, t_3, t_4$) within a driver or OS is challenging due to the inherent and random transmit delay and receive delay, {\em i.e.}, $J^{out}$ and $J^{in}$ in Figure~\ref{fig:ptp}, respectively. 
Since KeepON enables the deterministic packet scheduling function, we are able to mitigate the uncertainty in TX timestamping by calculating the outgoing EPHC time. 
For RX timestamps, we follow the software timestamping approach where receive delays are inevitable. 

\begin{figure}[tb]
    \centering
    \includegraphics[width=0.45\textwidth]{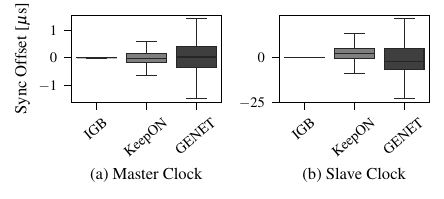}
    \vspace{-0.15in}
    \caption{PTP synchronization accuracy comparison among KeepON, GENET, and IGB in a single-hop network.}
    \label{fig:ptp_compare}
    \vspace{-0.15in}
\end{figure}

\vspace{0.02in}
\noindent \textbf{Synchronization accuracy measurements.} 
We compare the PTP synchronization accuracy among KeepON, GENET, and IGB in a single-hop network with slot size and batch size in DMA set to 300 bytes and 1, respectively. As shown in Figure~\ref{fig:ptp_compare}, hardware timestamping achieves the highest accuracy as expected (around 0.015\,$\mu$s act as both master and slave). On the other hand, KeepON demonstrates significantly improved synchronization accuracy (0.3\,$\mu$s as master and 3.7\,$\mu$s as slave) over software timestamping (0.5\,$\mu$s as master and 6.9\,$\mu$s as slave), due to more precise TX timestamps. To better understand the synchronization offset that arises from inaccurate TX/RX timestamping, we also conduct a theoretical synchronization analysis (detailed in Appendix~\S\ref{ssec:ptp-analys}), which shows that both clock granularity and timestamping jitter linearly affect synchronization accuracy. 

\subsection{Dual Clock Mechanism}

To address the inherent inaccuracy of receive timestamps, we develop a Dual Clock Mechanism (optional) that maintains two distinct EPHC clocks simultaneously. 
We establish a second RX-EPHC, driven by a packet counter tracking incoming packets on the receive path. This requires the remote peer device also to transmit packets continuously at the line rate, to ensure a stable stream of incoming packets to drive the RX counter. 
The receive path utilizes another CP-PMD polling loop dedicated to monitoring the RX DMA ring. 

\vspace{0.02in}
\noindent \textbf{Clock merging.} The proposed dual EPHC setup provides two independent clocks and more accurate timestamping mechanism for both transmit and receive paths. However, the system requires a unified time reference for overall consistent operation and PTP synchronization. Therefore, we propose a clock merging algorithm to derive a unified clock time by dynamically merging the instantaneous values from the TX-EPHC $T_{tx}$ and RX-EPHC $T_{rx}$. 
The algorithm handles two cases:

\vspace{0.02in}
\noindent $\bullet$ \textit{Large divergence:} If the absolute difference $|T_{tx} - T_{rx}|$ exceeds a predefined $2\tau$ worst-case clock jitter, we set the lagging clock to the leading clock. This is based on our practical observation of PMD that if the two clocks diverge significantly, it indicates a potential fault in the slower clock due to traffic burst or system jitter. 

\vspace{0.02in}
\noindent $\bullet$ \textit{Small divergence:}  If the clocks are consistent with $|T_{tx} - T_{rx}| \leq 2\tau$, their values are combined using a heuristic function to produce a merged timestamp. Under a non-faulty state, we observe that clock jitter is predominantly due to delayed clock updates. Thus, we implement the merging function as $[max(T_{tx},T_{rx}) + min(T_{tx} + \tau, T_{rx} + \tau)] / 2$. Then, assuming $T_{tx}$ and $T_{rx}$ have equal clock performance, the maximum possible jitter of the merged timestamp is reduced from $\tau$ to $\tau/2$.

\begin{figure}[tb]
    \centering
    \includegraphics[width=0.45\textwidth]{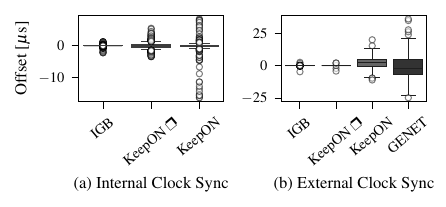}
    \vspace{-0.15in}
    \caption{Comparison of clock stability and synchronization accuracy for single clock (KeepON), dual clock (KeepON \ding{112}), software clock (GENET) and hardware clock (IGB).}
    \label{fig:ptp_dual}
    \vspace{-0.15in}
\end{figure}

\vspace{0.02in}
\noindent \textbf{Experimental validation.} 
Experimental results show that the dual clock mechanism not only significantly improves synchronization accuracy but also improves clock stability. %
Figure~\ref{fig:ptp_dual}(a) compares the clock stability, where the dual clock mechanism (KeepON~\ding{112}) achieves slightly better stability with an absolute offset of 0.58\,$\mu$s, compared to the single clock mechanism (KeepON) with an absolute offset of 0.64\,$\mu$s. 
More importantly, KeepON~\ding{112} has a much better 99th value of 5.3\,$\mu$s, compared to that of KeepON (16.4\,$\mu$s). This is because KeepON~\ding{112} tends to merge two clocks with a large divergence to correct potential faults. 
Figure~\ref{fig:ptp_dual}(b) shows the synchronization accuracy by comparing the offset with external clock. KeepON~\ding{112} achieves higher accuracy with an absolute offset of 10\,ns, compared to KeepON with 3.8\,$\mu$s. Moreover, KeepON~\ding{112} is shown to be slightly better than the hardware clock with an average absolute offset of 15\,ns. 

%% file: 5-modules.tex
\section{Heterogeneous Traffic Management}\label{sec:traffic}
To support heterogeneous industrial traffic, the designed traffic management module aims to isolate real-time traffic and best-effort traffic through DMA ring partitioning. 

\subsection{Traffic Isolation}\label{ssec:isolation}
\begin{figure}[tb]
    \centering
    \includegraphics[width=0.45\textwidth]{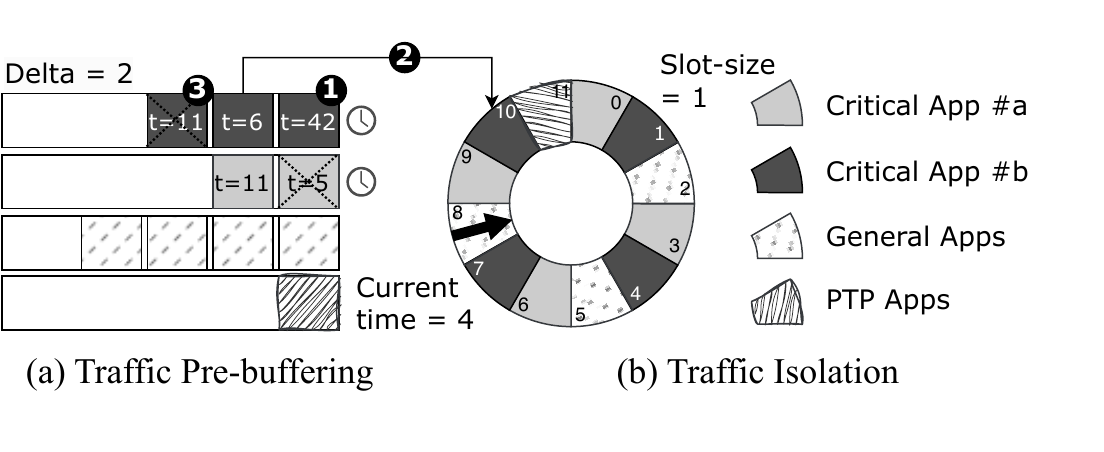}
    \vspace{-0.15in}
    \caption{An illustration of traffic pre-buffering and traffic isolation mechanisms, including partitions for two real-time traffic classes (critical App\#a and App\#b), and two best-efforts traffic classes (general services and PTP).}
    \label{fig:heto}
    \vspace{-0.15in}
\end{figure}

To achieve traffic isolation, we partition the DMA ring buffer into dedicated regions, with each region assigned to a specific traffic class. For example, as shown in Figure~\ref{fig:heto}(b), the DMA ring buffer is partitioned into four distinct regions, each serving either a critical application (e.g., App \#a and App \#b) or best-effort traffic. This partitioning ensures strict isolation, if the packet of an application intends to be inserted into a slot in the ring buffer that is reserved for another traffic class ({\em e.g.}, App \#a inserts t=11, belongs to App \#b \ding{204}), the packet will be dropped. 

Based on the partitioning mechanism, we need to determine the partition allocated to each traffic class to ensure that the resources specified by the partition can meet the timing requirements of the critical application. Unlike traditional time-triggered scheduling, our problem uniquely maps continuous-time transmissions onto a circular buffer where a packet scheduled at time t must occupy the slot determined by slot-size and ring size, and transmissions at different time compete for the same physical slot. To achieve this, we model this as a resource partitioning and traffic scheduling problem and provide the definition of a feasible schedule guaranteeing their timing requirements. The detailed modeling and formal constraints for this partitioning problem are provided in~\S\ref{ssec:isolation_appendix}.

We employ Satisfiability Modulo Theories (SMT)~\cite{barrett2021satisfiability} to translate the above variables and constraints into a system of logical and arithmetic constraints. An SMT solver, such as Z3~\cite{de2008z3}, can systematically search for a solution that satisfies all these constraints. If a solution is found, it readily provides a feasible partition and schedule; conversely, if the solver returns unsatisfiability, it indicates that no available partition exists under the given parameters. In this case, one can either increase the DMA ring size or change the efficiency parameters of PMD, {\em e.g.}, reducing the slot size.

\subsection{Traffic Pre-Buffering}
\label{ssec:prebuffer}
If applications submit packets too early before the scheduled times that fall out of the range of the immediate insertion window discussed in~\S\ref{sec:scheduled-packet-insertion}, the packets need to be pre-buffered until the available window. 
To address this, we introduce the \textit{traffic pre-buffering} mechanism. This involves maintaining intermediate software queues that decouple packet submission from immediate DMA insertion, allowing for orderly management and timely hand-off based on traffic class. Figure~\ref{fig:heto}(a) illustrates this concept, where packets are pre-buffered in software queues before being inserted into the DMA ring buffer.

\vspace{0.02in}
\noindent $\bullet$ \textit{Real-time traffic} with application-defined target transmission times ($t_{sch}$), are managed in a min-heap pre-buffering queue ordered by their scheduled timestamps~\cite{tc-etf}. A separate software component periodically checks the heap's root packet (earliest schedule) against the EPHC's current time. If the root packet's $t_{sch}$ enters immediate insertion window, it's removed from the heap and sent to the PMD for scheduled insertion. The min-heap ensures real-time packets are inserted in earliest-deadline-first order.

\vspace{0.02in}
\noindent $\bullet$ \textit{Best-effort traffic} without precise deadlines, uses a simpler FIFO queue for pre-buffering, with packets enqueued as received. When the Traffic Isolation logic identifies an available slot in the designated DMA ring partition ({\em e.g.}, a placeholder slot at near future), a packet is dequeued from the FIFO's head. This packet is then sent to PMD to be placed into the next available placeholder slot within its assigned partition.

%% file: 6-evaluation.tex
\section{Microbenchmarks}\label{sec:evaluation}
We evaluate the performance of KeepON across two complementary scenarios: 1) controlled microbenchmarks on a point-to-point link with two end devices directly connected via Ethernet to quantify KeepON's core metrics (\S\ref{sec:evaluation}); and 2) a real-world case study on a multi-hop TSN testbed with switches to demonstrate end-to-end effectiveness (\S\ref{sec:case-study}).

In this section, we focus on the microbenchmark results, assessing KeepON's performance at the end devices by measuring TX jitter/latency, bandwidth efficiency, CPU utilization, and energy efficiency, in comparison to the default driver and hardware-offloading baselines. KeepON is implemented atop the GENET driver for the BCM2711 SoC's Ethernet controller (see Appendix~\S\ref{ssec:driver-impl} for implementation details).

\subsection{Experiment Setup}
Two end devices are directly connected via an Ethernet cable. The sender runs on either: 1) a Raspberry Pi 4 Model B (Cortex-A72 @ 1.5GHz, 8GB LPDDR4 RAM) with its onboard Ethernet adapter for testing both the original GENET driver and KeepON; 2) a Raspberry Pi CM4 with PCIe I/O board equipped with an Intel i210 NIC running the IGB driver. We configure Linux ETF Qdisc differently for each approach: KeepON/IGB use it solely for pre-buffering due to their scheduling capabilities, while GENET relies on ETF for software-based timing control. We use default driver settings for GENET and IGB unless otherwise specified. All experiments operate at 1\,Gbps line rate. The receiver consistently uses an Intel i210 NIC with IEEE 1588 PTP hardware timestamping (8\,ns resolution) to measure jitter and delay at the MAC layer, excluding receiver network stack effects from our analysis.

\begin{figure}[tb]
    \centering
    \includegraphics[width=0.45\textwidth]{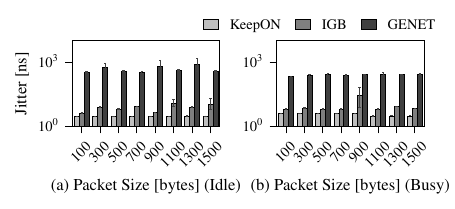}
    \vspace{-0.15in}
    \caption{TX jitter comparison with varied packet sizes and background traffic conditions.} %
    \label{fig:final_rt_single}
    \vspace{-0.15in}
\end{figure}

\subsection{TX Determinism}
We evaluate TX determinism by measuring the inter-arrival times of packets at the receiver side over a back-to-back Ethernet connection, which directly reflects the sender’s transmission timing accuracy. We compare KeepON with GENET and IGB under two background traffic conditions, idle and busy, while varying the packet size of real-time traffic. In each experiment, we transmit $1 \times 10^6$ packets with a target period of 1\,ms, where the application issues packets 100\,$\mu$s ahead of their scheduled transmission time.

Figure~\ref{fig:final_rt_single}(a) presents the results under idle background traffic. KeepON consistently achieves the lowest inter-arrival jitter, recording 3.7\,ns, compared to 7.7\,ns for IGB and 480.1\,ns for GENET. Under busy background traffic (Figure~\ref{fig:final_rt_single}(b)), we observe similar trends, confirming that KeepON provides the most accurate scheduling across varying stream sizes and traffic conditions. KeepON outperforms IGB because it schedules transmissions by aligning them to the line-rate byte flow at byte-level (8\,ns) granularity, whereas the i210's LaunchTime logic operates at a coarser 32\,ns granularity~\cite{intel-i210-datasheet}.

\subsection{TX Latency}
We further evaluate transmission path latency by examining how close to the scheduled transmission time an application can generate data while still maintaining low jitter. A shorter transmission path latency enables fresher data delivery, thereby reducing data aging. To assess this, we vary the advance time (5–50\,$\mu$s) and the traffic period (20–100\,$\mu$s), and measure receiver-side jitter following the same methodology as in the previous jitter analysis. This indirect approach avoids the timestamp synchronization challenges that complicate direct delay measurements.

\begin{figure}[tb]
    \centering
    \includegraphics[width=0.45\textwidth]{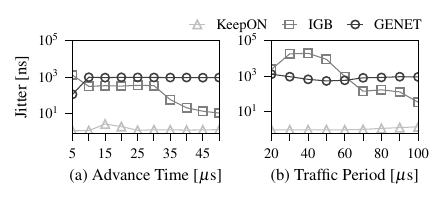}
    \vspace{-0.15in}
    \caption{TX latency comparison under varying advance times and traffic periods.} \label{fig:tx_delay_comparison}
    \vspace{-0.15in}
\end{figure}

Figure~\ref{fig:tx_delay_comparison}(a) shows that, for a 1\,ms traffic period, KeepON maintains low jitter ($\leq$ 10\,ns) even with only 5\,$\mu$s advance time, whereas IGB requires at least 30\,$\mu$s to achieve stable operation and still exhibits much higher jitter ($\geq$ 0.3\,$\mu$s). This demonstrates KeepON’s ability to accept data closer to transmission deadlines. Figure~\ref{fig:tx_delay_comparison}(b) evaluates varied traffic periods with a fixed 100\,$\mu$s advance time. KeepON successfully handles 20\,$\mu$s periods with jitter below 10\,ns, while IGB fails to sustain reliable operation at any period $\leq$ 100\,$\mu$s. These results highlight KeepON’s efficient interrupt-free PMD operations and lightweight scheduling mechanism, enabling tighter timing guarantees than hardware-offloading solutions.

\subsection{Bandwidth Efficiency}
\label{ssec:bw}

\vspace{0.02in}
\noindent \textbf{Packet padding.}
This experiment quantifies the bandwidth overhead incurred by fixed-size transmission slots when packet payloads are smaller than the allocated slot size. The overhead arises only when the slot size exceeds the payload length, whereas perfectly matching the slot size to the payload eliminates any overhead.

\begin{figure}[tb]
    \centering
    \includegraphics[width=0.45\textwidth]{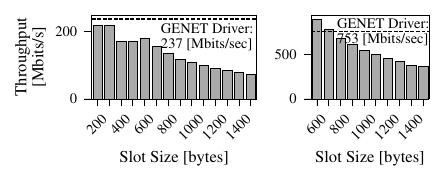}
    \vspace{-0.15in}
    \caption{Evaluation of KeepON's fixed slot size mechanism and the impact of padding on UDP throughput.}
    \label{fig:bandwidth-padding}
    \vspace{-0.2in}
\end{figure}

Figure~\ref{fig:bandwidth-padding} reports bandwidth measurements for 100-byte and 500-byte payloads under varying UDP slot sizes. For 100-byte payloads (Figure~\ref{fig:bandwidth-padding}(a)), throughput peaks at 218\,Mbps with 200–300 byte slots but drops to 73.3\,Mbps with 1500-byte slots due to excessive padding. For 500-byte payloads (Figure~\ref{fig:bandwidth-padding}(b)), 600-byte slots achieve nearly 895\,Mbps, whereas 1500-byte slots deliver only 367\,Mbps. These results indicate that selecting the smallest slot size that accommodates the typical payload maximizes efficiency. This is particularly relevant for control systems, where packet sizes are usually uniform ({\em e.g.}, our TSN case study pairs 200-byte packets with 200-byte slots). Larger slot sizes are preferable only when reducing CPU overhead and packets-per-second (PPS) handling is prioritized over payload efficiency (\S\ref{ssec:cpu}).

\vspace{0.02in}
\noindent \textbf{Traffic isolation.}
This experiment evaluates the effectiveness of the traffic isolation mechanism by measuring best-effort (BE) traffic bandwidth under different slot allocation settings. We use \texttt{iPerf2} to generate BE traffic and vary the reserved slots as follows: KeepON-25\% (8/32 slots), KeepON-50\% (16/32 slots), and KeepON-100\% (32/32 slots). Tests are conducted with UDP packets of 200, 400, 600, and 1400 bytes. Each run lasts 60 seconds and is repeated 10 times, with results averaged across repetitions.

\begin{figure*}[tbp]
    \centering
    \includegraphics[width=.95\textwidth]{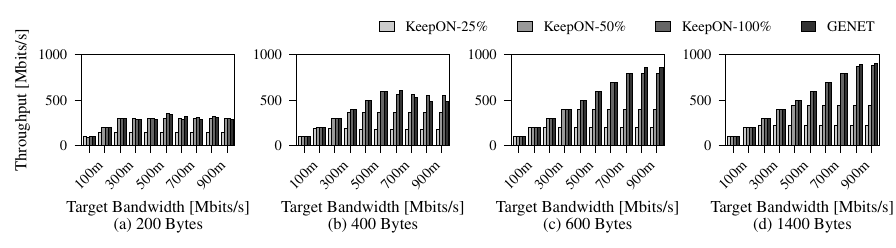}
    \vspace{-0.15in}
    \caption{Best-effort traffic throughput under KeepON's traffic isolation mechanism with varying UDP packet sizes.}
    \label{fig:bandwidth-slots}
    \vspace{-0.15in}
\end{figure*}

Figure~\ref{fig:bandwidth-slots} demonstrates that KeepON effectively regulates BE traffic throughput via traffic isolation, providing predictable bandwidth control. With KeepON-100\%, BE throughput closely matches that of GENET across all packet sizes and bandwidth settings. For 1400-byte packets, both GENET and KeepON-100\% approach the target bandwidth of 800\,Mbps before reaching saturation. For smaller packets (200 bytes), lower saturation points are observed due to higher per-packet processing overhead on a single CPU core. Notably, KeepON-100\% occasionally outperforms GENET under high loads (above 600\,Mbps), benefiting from PMD’s more efficient processing compared to interrupt-driven handling.

These results demonstrate that KeepON provides precise isolation without incurring additional bandwidth overhead. Best-effort throughput closely follows the configured slot share (25\%, 50\%, 100\%) across different packet sizes, confirming accurate enforcement of isolation. When provisioned at 100\%, KeepON matches and occasionally exceeds the baseline GENET driver, showing that best-effort performance is not compromised when full resources are available.

\subsection{CPU Utilization}
\label{ssec:cpu}

\begin{figure}[tb]
    \centering
    \includegraphics[width=0.45\textwidth]{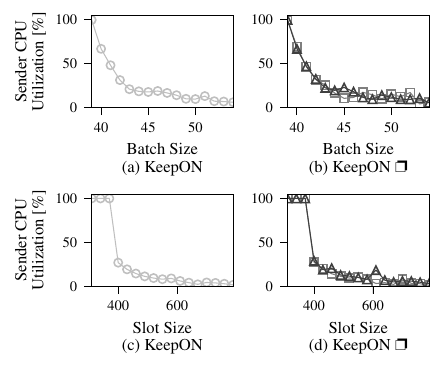}
    \vspace{-0.15in}
    \caption{Sender CPU utilization of KeepON's CP-PMD as a function of batch size and slot size. CPU utilization was measured using mpstat with the system under idle state.}
    \vspace{-0.15in}
    \label{fig:cpu}
\end{figure}

The PMD operation in KeepON consumes CPU resources to support periodic ring reclamation and line-rate transmission. Figure~\ref{fig:cpu} reports CPU utilization of PMD cores, measured with \textit{mpstat}, under varying batch sizes and slot sizes in both single-clock and dual-clock configurations. Each measurement represents the average utilization over a 60-second run with idle background traffic. When varying batch size, the slot size is fixed at 300 bytes; when varying slot size, the batch size is fixed at 1.

\vspace{0.02in}
\noindent \textbf{Batch size.} CPU utilization is highly sensitive to batch size in both configurations. In single-clock mode (Figure~\ref{fig:cpu}(a)), utilization remains near 100\% for batch sizes below 39, then drops sharply as the batch size increases, reaching 6.3\% at a batch size of 54. Dual-clock mode (Figure~\ref{fig:cpu}(b)) exhibits similar behavior across both TX and RX cores, with utilization decreasing from 100\% to approximately 6\% at batch size 54.

\vspace{0.02in}
\noindent \textbf{Slot size.} CPU utilization is strongly influenced by slot size in both configurations. In single-clock mode (Figure~\ref{fig:cpu}(c)), PMD core utilization remains at 100\% for slot sizes $\leq$ 370 bytes, decreases to 26.9\% at 400 bytes, and stabilizes around 1.7\% for 790-byte slots. Dual-clock mode (Figure~\ref{fig:cpu}(d)) exhibits the same trend across both TX and RX cores.

These results indicate that PMD CPU overhead is highly sensitive to batch and slot size. Proper tuning can reduce utilization to minimal levels while sustaining full performance.

\subsection{Energy Efficiency}
\label{ssec:energy}
We evaluate KeepON's energy efficiency by measuring power consumption under various experimental settings. The baseline driver GENET exhibits load-dependent power consumption, starting at 1.9W at idle and increasing linearly with traffic load to 4.6W at 1\,Gbps throughput. In contrast, KeepON's power profile depends significantly on PMD configuration, revealing two distinct operational regimes: \textit{1) Non-optimized configs}: KeepON consumes 3.4W at idle, which is 1.5W higher than GENET but maintains nearly constant power consumption regardless of bandwidth utilization; \textit{2) Optimized configurations}: With larger batch-size or slot size reduce KeepON's idle power to 2.7W (only 0.8W above GENET) while preserving the constant consumption profile. Crucially, this configuration makes KeepON more energy-efficient than GENET at traffic rates $\geq$300\,Mbps. For example, at 1\,Gbps, GENET consumes 4.6W while KeepON maintains 2.7W with a 41\% reduction in power consumption.

\begin{figure*}[tb]
    \centering
    \includegraphics[width=.98\textwidth]{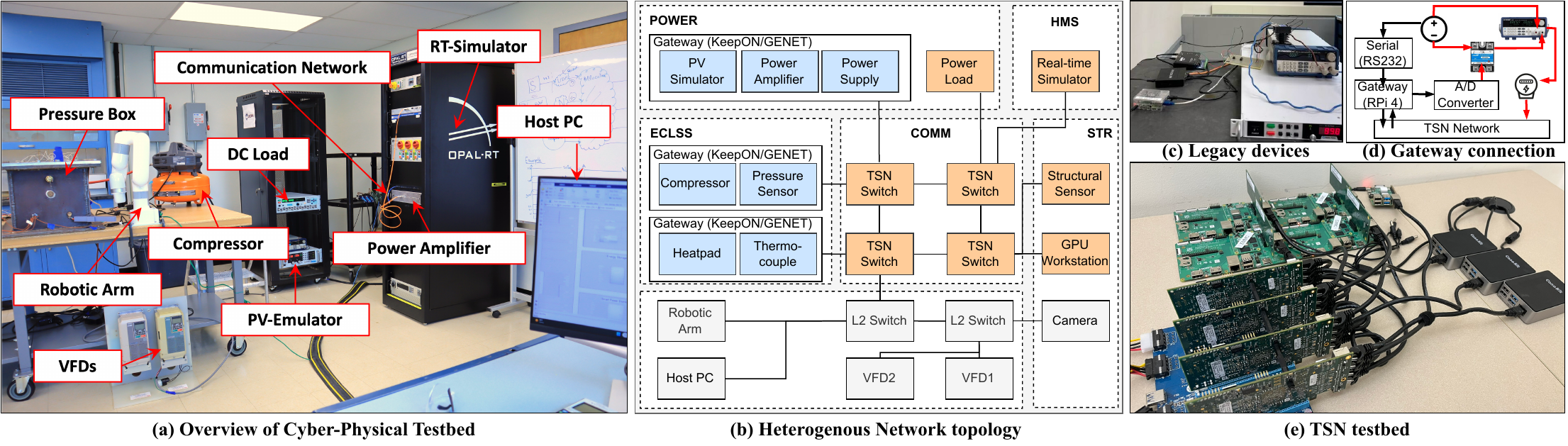}
    \caption{Overview of the RETHi Cyber-Physical Testbed (a-d) and TSN testbed (e).}
    \label{fig:testbed}
    \vspace{-0.15in}
\end{figure*}

%% file: 7-case.tex
\section{Case Study}
\label{sec:case-study}
We evaluate KeepON’s effectiveness through a case study on the Cyber-Physical Testbed (CPT) deployed at UConn, a multi-hop heterogeneous TSN network (Figure~\ref{fig:testbed}). Using the CPT’s network topology (Figure~\ref{fig:testbed}(b)) and traffic flows (Table~\ref{tab:testbed-flow-param}), we replicate the network backbone on a TSN testbed (Figure~\ref{fig:testbed}(e)) to conduct experiments. This setup allows direct capture of MAC-layer timestamps for accurate measurement of delay and jitter, while providing flexible adjustment of flow sampling rates for long-duration evaluations.

\vspace{0.02in}
\noindent \textbf{RETHi Cyber-Physical Testbed (CPT).} CPT supports deep-space habitat research by integrating hardware and simulation components over a 1\,Gbps TSN network (Figure~\ref{fig:testbed}(a,b)). Cyber components, {\em e.g.}, the modular space habitat model, nuclear power system, and environmental control and life support system (ECLSS), run on a real-time simulation platform ({\em e.g.}, Opal-RT). Physical components—including structural elements, power distribution and consumption systems, and a robotic arm—are implemented as hardware under test.

The CPT includes both TSN-capable and legacy components due to cost and operational constraints. High-performance real-time devices, such as Opal-RT, feature TSN NICs with IEEE 1588 PTP hardware clocks and scheduling offloading, enabling deterministic communication. In contrast, many physical components rely on legacy interfaces—for example, the power supply uses RS-232 and power relays receive analog signals (Figure~\ref{fig:testbed}(c,d)). To integrate these devices, we employ three Raspberry Pi 4B units as network gateways, which are cost-effective, easily updatable, and support multiple I/O interfaces (serial, GPIO, WiFi, Ethernet). Since the RPi 4B’s BCM2711 NIC lacks scheduled transmission support, we prototype KeepON on this platform and compare its performance against the original GENET Ethernet driver in TSN integration.

\vspace{0.02in}
\noindent \textbf{TSN Testbed.} Figure~\ref{fig:testbed}(e) shows our TSN testbed, which consists of four TTTech FPGA-based network bridges, three RPi 4B devices running KeepON to emulate legacy components, four RPi CM4 units with TSN NICs emulating real-time components, and a Central Network Controller (CNC) on an RPi 4B responsible for schedule computation and distribution. The RPi CM4 listeners use IEEE 1588 hardware timestamping to capture MAC-layer packet arrival times.

We evaluate 12 critical flows representative of RETHi CPT traffic patterns (details in Appendix~\S\ref{sec:appendix-flows}). Among these, flows F2–F4 and F7–F8 are generated by KeepON/GENET, while the remaining flows are handled via hardware scheduling offload from TSN NICs. All flows transmit uniform 200-byte packets with 1-ms periods, forming a 1-ms hyperperiod. The TSN schedule is generated using~\cite{pahlevan2019heuristic}. KeepON’s PMD slot size matches the packet size, and ring buffer isolation is configured according to each flow’s period and offset to ensure compliance with the global TSN schedule.

\vspace{0.02in}
\noindent \textbf{Results.} We measure schedule-to-delivery delay for seven critical flows traversing the multi-hop TSN testbed. Table~\ref{tab:flow-stats-delay} shows that KeepON maintains deterministic delays closely matching the expected TSN schedule, with sub-microsecond jitter. In contrast, under the same TSN schedule, GENET exhibits significantly higher delays, with P99 delays for flows F3, F4, and F7 exceeding their 1-ms period.

\input{figs/testbed/stats.tex}

%% file: figs/testbed/stats.tex
\begin{table}[t]
  \caption{Delay comparison across 7 critical flows.}
  \label{tab:flow-stats-delay}
  \small
  \centering
  \resizebox{0.95\columnwidth}{!}{
  \begin{tabular}{|c|c|c|c|c|c|c|}
    \hline
    \multirow{2}{*}{\textbf{FlowId}} 
      & \multicolumn{3}{c|}{\textbf{GENET Delay [$\mu$s]}} 
      & \multicolumn{3}{c|}{\textbf{KeepON Delay [$\mu$s]}} \\
    \cline{2-7}
      & {Mean} & {P99} & {Max} & {Mean} & {P99} & {Max} \\
    \hline
    1  &   8.8 &    8.8 &     8.8 &   8.8 &    8.8 &    12.7 \\
    \hline
    2  &  15.5 &   15.6 &    19.7 &  12.4 &   12.4 &    16.5 \\
    \hline
    3  &  29.2 & \textbf{1008.8} &  \textbf{1008.8} &  12.4 &   12.4 &    12.4 \\
    \hline
    4  & \textbf{936.7} & \textbf{1010.5} &  \textbf{1010.5} &  12.1 &   12.1 &    12.1 \\
    \hline
    7  & \textbf{980.6} & \textbf{1009.2} &  \textbf{1009.2} &  12.4 &   12.4 &    12.4 \\
    \hline
    8  &  10.3 &   10.5 &    12.4 &   8.5 &    8.5 &     8.6 \\
    \hline
    12 &   5.2 &    5.2 &     5.2 &   5.2 &    5.2 &     5.2 \\
    \hline
  \end{tabular}}
\vspace{-0.2in}
\end{table}

%% file: 8-related.tex
\section{Related Work}
Existing solutions for end devices to achieve deterministic communications fall into two categories (see Appendix~\S\ref{sec:related} for detailed discussion).

\vspace{0.01in}
\noindent\textbf{Hardware-dependent approaches} include FPGA/SoC-based systems~\cite{quan2020opentsn,kyriakakis2020time,li2023deterministic} that achieve precise timing but lack flexibility and require specialized platforms. More practical solutions widely use general systems with specialized NICs that support hardware offloading~\cite{coleman2019emerging,bosk2022methodology,grigorjew2022affordable,xue2024towards}, offering better integration, but still requiring specific hardware that is unavailable in legacy and embedded systems.

\vspace{0.01in}
\noindent\textbf{Hardware-independent approaches} focus on reducing average latency rather than providing TX determinism~\cite{saeed2019eiffel, hwang2015netvm, zhou2024dint, tu2021revisiting, sadok2023enso, alcoz2020sp, raghavan2023cornflakes}. While the kernel provides software TSN mechanisms like TAPRIO and ETF for traffic shaping and scheduling, their deterministic performance degrades significantly without hardware support~\cite{tc-etf, ulbricht2023tsn}. For specialized network stack targets determinism, pure software approaches still suffer from tens of microseconds of jitter, inherent random delays from the OS and PCIe, and fail the TSN requirements~\cite{scheddeadline, brinkman2021fastcat}.

\section{Conclusion}
\label{sec:conclusion}

This paper presents KeepON, a novel software-based network driver model supporting deterministic traffic on end devices equipped with standard NICs. 
KeepON's driver model integrates three key components, Continuous-Pacing PMD, Emulated PTP Hardware Clock, and Scheduled Packet Insertion, to achieve precise and deterministic traffic scheduling. To extend support for end-to-end determinism for network integration and heterogeneous traffic types, we further propose two additional modules: the Synchronization module and the Traffic Management module. 
Extensive evaluations of our prototype, including both microbenchmarks and testbed integration, demonstrate that KeepON outperforms standard drivers and achieves performance comparable to hardware-based solutions. These results mark a significant step toward enabling deterministic communication on commodity platforms, offering a more flexible and accessible path to deploying deterministic networking in real-world systems.

%% file: ack.tex
\section{Acknowledgement}

The work is supported in part by the National Science Foundation Grant CNS-2008463, CNS-2432533 and the NASA STRI Resilient Extraterrestrial Habitats Institute (RETHi) under grant number 80NSSC19K1076.

%% file: 9-appendix.tex
\section{Appendix}

\vspace{0.1in}
\subsection{Detailed Related Work}
\label{sec:related}

Existing works supporting deterministic communication on end devices can be broadly classified into hardware-dependent and hardware-independent solutions.

\vspace{0.02in}
\noindent\textbf{Hardware-Dependent Solutions.}
{Several hardware-based approaches have been developed to enhance network performance and scheduling efficiency by leveraging specialized hardware features. For instance, SENIC~\cite{radhakrishnan2014senic} demonstrates scalable NIC rate limiting on NetFPGA. Loom~\cite{stephens2019loom} provides flexible NIC packet scheduling through hardware modifications to improve scheduling efficiency. Azure Accelerated Networking~\cite{firestone2018azure} demonstrates the production-scale deployment of FPGAs, achieving low VM-VM latencies and 32\,Gbps throughput. Programmable Calendar Queues~\cite{sharma2020programmable} enable time-aware priority escalation on programmable switches, such as Barefoot Tofino, for algorithms requiring dynamic packet priority changes. While these approaches significantly improve performance and scheduling flexibility, they focus on throughput optimization and average latency reduction rather than providing strict timing determinism.
}

There are several hardware-dependent solutions that aim to achieve deterministic communication at end devices by leveraging specific hardware features, which provide highly precise scheduling and synchronization capabilities.

\vspace{0.02in}
\noindent $\bullet$ \textit{Specialized Platform.} One type of solution relies on FPGAs or specialized System-on-Chip (SoC).
For instance, Quan et al. \cite{quan2020opentsn} propose an SDN-based TSN control mechanism using Xilinx FPGA with a novel management protocol. Kyriakakis et al.~\cite{kyriakakis2020time} present the design and implementation of a time-predictable TTEthernet end system based on the open-source Patmos processor. In~\cite{li2023deterministic}, an open-source FPGA TSN end device implementation is proposed to support precise time-triggered task release. 
However, the specialized nature of these FPGA/SoC-based solutions limits their applicability and easy integration into existing systems.

\noindent $\bullet$ \textit{General Systems with Specialized NICs}. Another common and more widely adopted hardware-dependent approach utilizes general-purpose OS (e.g., Linux) augmented with specialized NICs that support hardware offloading.
In this thread of work, several research efforts have been conducted to improve delay, jitter, and system throughput~\cite{coleman2019emerging, bosk2022methodology, grigorjew2022affordable}. Besides interacting via standard Linux APIs, some other works employ user-space frameworks, e.g., DPDK, to bypass kernel overhead and gain finer control~\cite{nayak2016time, xue2024towards}. This approach is flexible since it leverages a standard OS and typically does not require extensive, ground-up development, making it cost-effective. 
However, it still requires specific NIC hardware, which is unavailable in legacy systems, and some embedded systems may not even have an available PCIe port to install such an extension NIC, such as Raspberry Pi and Arduino.

\vspace{0.02in}
\noindent\textbf{Hardware-Independent Design}
{Several existing studies aim to reduce packet transmission latency at the network stack or OS level using hardware-independent techniques, including kernel bypass~\cite{hwang2015netvm}, accelerated eBPF/XDP processing~\cite{zhou2024dint, tu2021revisiting}, and streamlined network stacks~\cite{sadok2023enso}. Recent software packet scheduling advances, such as Eiffel~\cite{saeed2019eiffel}, which uses integer-based priority queues to achieve $O$(1) scheduling operations, and SP-PIFO~\cite{alcoz2020sp}, which approximates complex scheduling on commodity hardware, demonstrate significant efficiency improvements. However, these approaches focus on packet ordering (priority) rather than precise transmission time. Cornflakes~\cite{raghavan2023cornflakes} demonstrates $\mu$s-scale networking via zero-copy serialization, achieving higher throughput than pure software approaches. Although these methods can reduce average latency, they do not guarantee strict timing determinism. 
}

Purely software-based approaches for deterministic communication remain rare and typically suffer from performance limitations compared to hardware solutions, largely due to inherent sources of random delay, as discussed in \S\ref{sec:intro}. Although the Linux kernel provides software TSN mechanisms like TAPRIO and ETF for traffic shaping and scheduling (operable even without hardware offloading), their deterministic performance degrades significantly without using with specialized NICs~\cite{tc-etf, ulbricht2023tsn}. The PREEMPT\_RT Linux kernel patches~\cite{realtimelinux} provide a foundation for kernel-level real-time networking, achieving ~100$\mu$s worst-case scheduling latency, while SCHED\_DEADLINE offers EDF scheduling with temporal isolation for time-critical tasks~\cite{scheddeadline}. Frühwirth et al. \cite{fruhwirth2015ttethernet} propose a software-based TTEthernet end system that achieves 30\,$\mu$s jitter for time-triggered transmissions. Similarly, Fastcat~\cite{brinkman2021fastcat}, an open-source C++ library for EtherCAT end devices using standard NICs, supports 1\,kHz real-time communication but lacks evaluation of its deterministic performance and remains limited to 100\,Mbps EtherCAT networks.

\subsection{Background}
\label{ssec:background_appendix}

\begin{figure}[tb]
    \centering
    \includegraphics[width=0.45\textwidth]{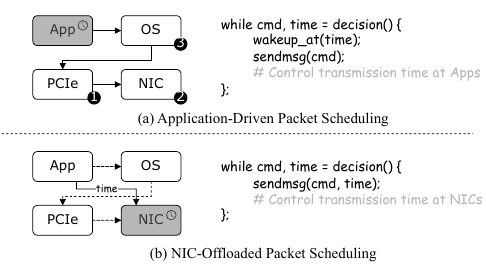}
    \caption{Comparison of packet transmission timing control methods on end devices.}%
    \label{fig:motivation_append}
\end{figure}

\noindent\textbf{Delay Components on End Devices.} In this section we discuss why scheduling offloading is necessary and simply reducing average delay cannot guarantee TX determinism. When a real-time flow is transmitted over a deterministic network, the data originates from the application layer on the end device, typically through a system call (e.g., \texttt{send\_msg}) that requests the operating system to dispatch the packet toward the network interface, as shown in Figure~\ref{fig:motivation_append}(a). The path of this data flow from the user space to the physical medium, however, includes a series of processing stages that collectively introduce non-deterministic delay. These delays originate from both operating system behavior and architectural constraints, and together they pose significant challenges to the deterministic timing performance of end systems. 

\vspace{0.02in}
\noindent $\bullet$ \textit{Architecture-level delay}. One major source of timing variability stems from contention on the PCIe fabric (\ding{202}), %
where multiple concurrent data transfers can cause unpredictable access latencies in the order of 10\,µs~\cite{neugebauer2018understanding, coleman2019emerging, xue2024towards}. 
Memory-mapped I/O (MMIO) operations (\ding{202}) may also experience fluctuating delays depending on the system load (e.g., worsened by small-write bursts) and cache coherency mechanisms~\cite{schuh2024cc, emmerich2018user}. 
Other sources of architecture-level delay include batching mechanisms used by modern PCIe and NICs (\ding{202}, \ding{203}), fetching policies implemented on NICs (\ding{203}), and latency introduced by I/O Memory Management Unit~\cite{IntelI225, nvidia, 8021qbv, bosk2022methodology, grigorjew2022affordable}.

\vspace{0.02in}
\noindent $\bullet$ \textit{OS-level delay} (\ding{204}). Further sources of non-determinism emerge due to the general-purpose nature of modern kernels, even when enabled with real-time capabilities. For example, dynamic resource management mechanisms, e.g., dynamic socket buffer tuning, introduce delay variability from buffer reallocation or page fault under memory pressure~\cite{madden2019challenges}. In addition, interrupts, context switches and task preemption in the network stack and scheduling process can both result in unpredictable delays~\cite{soares2010flexsc, xue2024towards}.

To deal with these non-deterministic timing variability, an increasingly adopted approach is to offload the responsibility of precise packet timing control from the OS (e.g., real-time scheduling, interrupt management) to the NIC~\cite{coleman2019emerging, bosk2022methodology, grigorjew2022affordable, nayak2016time, xue2024towards}. Examples include Intel's LaunchTime feature on i210/i225 controllers, NVIDIA Mellanox's Accurate Scheduling technology, and IEEE 802.1Qbv Time-Aware Shaper~\cite{IntelI225,nvidia,8021qbv}.
This approach fundamentally changes the timing control paradigm.
As shown in Figure~\ref{fig:motivation_append}(b), rather than relying on the CPU to initiate transmission at a specific time, the driver specifies a future transmission timestamp, and the NIC independently takes responsibility to enforce it. Once the timestamp is set, the packet resides in the NIC's hardware queue until the scheduled time when it is released onto the network medium. This strategy essentially bypasses all the previously discussed sources of non-deterministic delay. 

\noindent\textbf{Limited Support for Standard NICs.}
Supporting scheduling offloading functionality requires the NIC to have two distinct hardware capabilities: 1) a local onboard hardware clock with high-precision synchronization (typically via IEEE 1588 Precision Time Protocol), and 2) specific logic to schedule packet transmissions according to timestamps using that clock. 
These capabilities enable two primary offloading mechanisms in modern networking stacks. SO\_TXTIME is a Linux socket option that allows applications to specify the exact transmission time for individual packets, enabling the NIC to hold packets in its hardware queue and transmit them precisely at the designated timestamp. TAPRIO (Time-Aware Priority) implements the IEEE 802.1Qbv Time-Aware Shaper standard, which divides time into repeating cycles with scheduled time slots (gates) for different traffic classes, ensuring deterministic transmission windows for time-sensitive traffic.
However, after surveying 341 drivers from Linux kernel 6.12, we identify that only 46 drivers (13.5\%) support IEEE 1588 PTP hardware clock functionality, which provides the fundamental time synchronization required for any form of time-based scheduling. More critically, only 13 drivers (3.8\%) support hardware offloading for scheduled transmission through either SO\_TXTIME or TAPRIO mechanisms (see Table~\ref{tab:hardware_offloading}). 
Among these 13 drivers, the support is further fragmented: only 5 drivers support SO\_TXTIME for per-packet scheduling granularity, while 11 drivers support TAPRIO for time-slotted scheduling, with merely 3 drivers (Intel igb/igc and NXP enetc) supporting both mechanisms.
This severe limitation means that NICs with the requisite hardware support for time-based scheduling remain absent in most standard commodity networking equipment, creating a significant barrier to the widespread deployment of time-sensitive networking applications.

\begin{table*}[tb]
    \caption{13 out of 341 Ethernet drivers in Linux 6.14 has hardware support for scheduled transmission offloading.}
    \centering %
    \small %
    \begin{minipage}{\textwidth} %
    \begin{tabular*}{\textwidth}{@{\extracolsep{\fill}}lccccccccccc@{}} %
    \toprule
    \textbf{Driver} & \textbf{igb/igc} & \textbf{enetc} & \textbf{mlx4/5} & \textbf{bnxt} & \textbf{atlantic} & \textbf{cpsw} & \textbf{stmmac} & \textbf{rtsn} & \textbf{xilinx} & \textbf{engleder} & \textbf{lan966x} \\ \midrule
    Vendor          & Intel            & NXP            & Nvidia             & Broadcom      & Aquantia          & TI                 & STMicro         & Renesas       & Nvidia              & Engleder          & Microchip        \\
    SO\_TXTIME      & \checkmark              & \checkmark            & \checkmark                & \checkmark           & \checkmark               &-                &-             &-           &-                 &-               &-              \\
    TAPRIO          & \checkmark              & \checkmark            &-                &-           &-               & \checkmark                & \checkmark             & \checkmark           & \checkmark                 & \checkmark               & \checkmark              \\ \bottomrule
    \end{tabular*}
    \end{minipage} %
    \label{tab:hardware_offloading}
\end{table*}

\subsection{Driver Implementation}
\label{ssec:driver-impl}

\vspace{0.02in}
\noindent \textbf{DMA Ring Initialization.} 
We initialize the driver by pre-allocating static packet buffers (skb) using \texttt{dev\_alloc\_skb} for all descriptors in the transmit DMA ring. These buffers all have the same fixed size (by the slot size parameter) and are initialized with corrupted CRC values. We create the DMA mappings for these buffers using the streaming DMA API (\texttt{dma\_map\_single}) at startup. While streaming mappings are typically transient, we retain these mappings for the lifetime of the driver.
This requires explicit synchronization before the NIC accesses the data (after CPU writes), we choose this over the \texttt{dma\_alloc\_coherent} API to have finer control and avoid cache overhead associated with coherent buffers. 
Each descriptor has flags added to track its status and ownership (belonging to which traffic class). 
Initially, the NIC's consumer/completion index is set to 0, and the driver's producer index is set to batch size. 
We disable NIC interrupts and the NAPI subsystem for TX, which is standard practice for polling-mode drivers (PMDs) to minimize PCIe transaction overhead and processing latency. Finally, a dedicated polling thread is created for transmit operations using \texttt{kthread\_create\_on\_cpu}, ensuring it runs on a specific CPU core for minimum preemption.

\vspace{0.02in}
\noindent \textbf{Identify Transmitted Packets.} 
Two approaches are generally available to determine when the NIC has finished transmitting packets. 1) The driver periodically reads a NIC register (via MMIO), which indicates how many descriptors the NIC has processed (e.g., a completion or consumer index). 
The driver compares this NIC index with its own record of submitted packets (its producer index) to identify completed transmissions. This method is simple and compatible with most standard NICs. 
2) Some NICs update status directly within the descriptor memory itself, e.g., Intel i210/i225 using Descriptor Done (DD) bits, or others using Completion Queue Entries (CQEs)~\cite{intel-i210-datasheet, IntelI225}. The driver can poll this memory location, which involves less overhead compared to MMIO register reads, to check for completion flags. This can be more efficient, but requires specific NIC hardware support. 
In our implementation, we select the first approach (i.e., Index Comparison) for broader compatibility. Once the PMD identifies that a packet has been transmitted, we intentionally modify its CRC field in the buffer, setting it to \texttt{0x00} as corrupted, assuming a low probability of collision with any valid CRC needed. We use solution 1) in our prototype.

\vspace{0.02in}
\noindent \textbf{Polling Period.} 
Our polling thread reclaims transmitted buffer descriptors in batches, controlled by the batch size parameter. The polling frequency (period) is critical for achieving stable line-rate performance without overwhelming the CPU. It must be fast enough to free up descriptors before the transmit ring fills. The polling period is determined by the packet transmission time (depending on PMD's slot size) and the number of packets processed per poll (batch size). 
Specifically, the polling period must satisfy the following relationship to keep pace with the line rate: polling period $\leq$ line-rate / (slot size $\times$ batch size) - polling overhead. 
Here, polling overhead represents the worst-case time the driver takes to process one batch of completed packets. Table~\ref{tab:dense_tradeoff_tabularstar} validates parameter combinations (slot size and batch size) by measurements that satisfy this condition in the above equation and enable line-rate throughput. In practice, the polling thread uses \texttt{usleep\_range(lb, ub)} to control polling period, yielding the CPU. We set the sleep range based on the calculated maximum polling period and use a pessimistic overhead value of 100\,$\mu$s in our implementation.

\begin{table}[tb]
    \centering %
    \small
    \caption{Relative Throughput (\%) under varying batch size and polling period ($\mu$s) settings. Packet size is 300 bytes.}
    \label{tab:dense_tradeoff_tabularstar}
    \begin{tabular*}{\dimexpr 0.5\textwidth-2\tabcolsep\relax}{@{} l @{\extracolsep{\fill}} c c c c @{}}
        \toprule
        Batch size & \multicolumn{4}{c}{Polling-period [$\mu$s]} \\
        \cmidrule(lr){2-5}
         & 0 & 10 & 100 & 1000 \\
        \midrule
        1   & \ul{99.99\%} & 16.77\% & 2.18\%  & 0.00\%  \\
        8   & 100.00\% & \ul{100.00\%} & 18.31\% & 1.89\%  \\
        64  & 100.00\% & 100.00\% & \ul{100.00\%} & 15.30\% \\
        512 & 100.00\% & 100.00\% & 100.00\% & \ul{100.00\%} \\
        \bottomrule
\end{tabular*}
\end{table}

\vspace{0.02in}
\noindent \textbf{Packet Insertion.} 
For packet insertion, the driver first obtains the target TX ring descriptor index from the scheduler. The scheduler implements different insertion strategies based on traffic type:

{\vspace{0.01in}
\noindent $\bullet$ \textit{Real-time Traffic:} For packets with \texttt{SO\_TXTIME} timestamps, the scheduler calculates the target slot according to the scheduled transmission time. If the target slot is available and the packet arrived on time, it is placed in the target slot. Otherwise, the driver implementation supports two modes: (i) \textbf{Strict Mode} drops the packet to maintain hard real-time guarantees, and (ii) \textbf{Relaxed Mode} (used only when isolation is not aligned with schedule) finds the earliest available slot within the traffic class's allocated slots, trading timing precision for higher delivery rates. The slot index is calculated based on the expected scheduled time as \mbox{(($t_{sch}$ $-$ offset) $/$ cycle-period) $\%$ $N$}, where $N$ is the size of DMA ring. 

\vspace{0.01in}
\noindent $\bullet$ \textit{Best-effort Traffic:} Packets without timing constraints use any available slot not reserved for real-time traffic classes, ensuring they only utilize unreserved bandwidth.
}

After obtaining a valid slot index, the driver copies (\texttt{memcpy}) the packet data into a pre-allocated buffer on the transmit ring. 
A \texttt{dma\_sync\_single\_for\_device} call ensures the NIC observes the data by flushing cache, and the descriptor is marked as ready by setting its state flag. 
For PTP packets, the driver passes the scheduled timestamp to the kernel using \texttt{skb\_tstamp\_tx} for hardware timestamping. Since packet insertion operates in parallel with the polling mechanism, a spinlock protects the shared ring buffer from concurrent access.
It is worth noting that we implement a `Lazy Doorbell' mechanism. The expensive MMIO doorbell write to the NIC's producer index is batched and moved from `xmit' to CP-PMD at the end of every poll cycle.

\vspace{0.02in}
\noindent \textbf{Dual Clock Mode.} 
In dual-clock mode, Continuous-Pacing PMD creates another separate polling-based thread for the RX path to avoid the interrupt overhead. Compared with the TX polling mechanism, there is a timestamping problem on the RX side, where the NIC drops corrupted packets directly in hardware (as discussed in Section~\ref{sec:observation}). This means the software driver, operating in the polling RX thread, only sees the genuine packets arriving in memory (i.e., the DMA ring). Because the dropped packets are invisible to the driver, it loses the original sequence of arrivals. 
While a hardware counter (e.g., bcmgenet's \texttt{DMA\_P\_INDEX\_DISCARD\_CNT}) might report the number of dropped packets, the driver doesn't know when they were dropped relative to the packets it received. This prevents accurate timestamping of the receiving packet as it initially hits the NIC. There are two potential solutions: 1) Disable the NIC's hardware CRC check, forcing it to pass all packets (corrupted and genuine) to the driver, thus preserving the sequence for timestamping; 2) Use a smaller batch size when polling and timestamping, which could offer relatively improved accuracy. We pick the first one for dual-clock mode.

\begin{figure}[tb]
    \centering
    \includegraphics[width=0.4\textwidth]{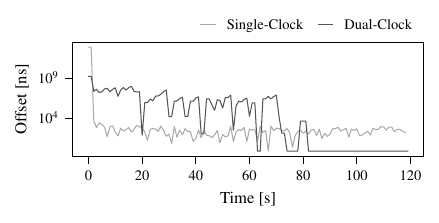}
    \caption{Convergence time of PTP synchronization of KeepON under Single-Clock vs. Dual-Clock mode.}
    \label{fig:converge}
\end{figure}

We identify two threats to the validity of the implementation of dual clock mode. 
1) Even though the dual clock mode can achieve accurate time synchronization after it converges, the convergence time is unpredictable and can be up to several minutes. For example, Figure~\ref{fig:converge} shows a case where the single clock mode is synchronized within 5 seconds, but the dual clock mode takes more than 80 seconds. We believe this is due to the algorithm of the dual clock combination used to merge the TX and RX clocks. 
2) We noticed that if we delay adding necessary padding data (discussed in Section~\ref{sec:scheduled-packet-insertion}) in reclaiming steps during PMD polling, it becomes harder to hit the line-rate target in dual-clock mode. This might be caused by cache pollution, where the two separate threads access the memory in an interleaved manner, leading to cache misses. Thus, for dual-clock mode, we add the padding on demand when it just reaches the driver.

\subsection{Supplementary Results}
\label{ssec:exp_appendix}

\begin{figure}[h]
    \centering
    \includegraphics[width=0.45\textwidth]{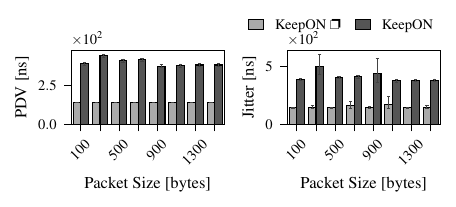}
    \caption{Deterministic performance comparisons of KeepON under single clock (KeepON) and dual clock (KeepON \ding{112}) modes, measured by software timestamping.}
    \label{fig:final_rt_dual}
\end{figure}

\vspace{0.02in}
\noindent $\bullet$ \textit{Dual Clock.}  Figure~\ref{fig:final_rt_dual} shows the deterministic performance comparison of single clock (KeepON) and dual clock (KeepON~\ding{112}) modes by varying packet size\footnote{KeepON's dual-clock mode requires both devices to operate with the KeepON driver model. This means that hardware timestamping, which is used for PDV comparisons in the single-clock mode (Figure~\ref{fig:final_rt_single}), cannot be applied here and other solutions are not included (e.g., GENET and IGB); thus, software timestamping is used in Figure~\ref{fig:final_rt_dual} for consistent comparison between KeepON's two modes.}.
The results show that the KeepON~\ding{112} further improves determinism, achieving consistently low PDV (packet delay variation) and inter-arrival jitter, averaging 141\,ns and 150\,ns, respectively, compared to 381\,ns and 416\,ns in the single-clock mode. These results clearly demonstrate that clock synchronization between KeepON-enabled devices in dual-clock mode substantially enhances deterministic network performance. 

\begin{figure}[h]
    \centering
    \includegraphics[width=0.45\textwidth]{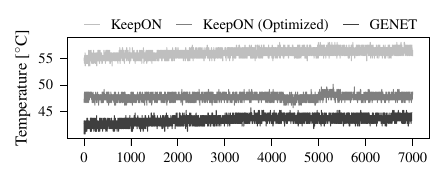}
    \caption{CPU thermal status comparisons.}
    \label{fig:thermal}
\end{figure}

\noindent\textbf{Thermal Efficiency.} We also compared the thermal impact of KeepON and the original GENET driver. Figure~\ref{fig:thermal} shows CPU temperatures over time for GENET, a high-stress PMD KeepON configuration, and an optimized KeepON setup. The unoptimized KeepON (slot size = 300 byte, batch size = 1) stabilized CPU temperatures at 54-56°C, significantly higher than GENET's 43-45°C, illustrating the thermal cost of aggressive polling for determinism. However, optimizing PMD settings (slot size = 1500, batch size = 32) in KeepON (Optimized) improved thermal efficiency, maintaining a lower 47-49°C. These results show that carefully tuned PMD parameters allow KeepON's energy efficiency to approach GENET's while preserving deterministic communication.

\begin{figure}[h]
    \centering
    \includegraphics[width=0.45\textwidth]{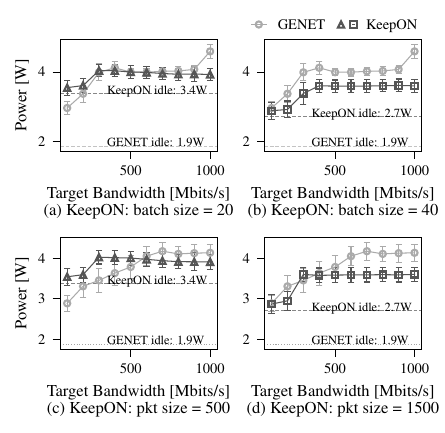}
    \vspace{-0.15in}
    \caption{Power consumption comparisons of KeepON and GENET under different settings.}
    \label{fig:power}
    \vspace{-0.15in}
\end{figure}

\vspace{0.02in}
\noindent\textbf{Energy Efficiency.} 
We evaluate the energy efficiency of KeepON by monitoring its power consumption under different experimental settings. As shown in Figure~\ref{fig:power}, the original GENET driver exhibits a lower idle power consumption of approximately 1.9~W, which then scales with the traffic load, reaching up to 4.6~W at 1\,Gbps. In contrast, KeepON's power profile is highly dependent on the polling parameter settings. Under non-optimized settings (Figure~\ref{fig:power}(a)(c)), such as a small batch size of 20 or a small packet size of 500 bytes, KeepON exhibits a significantly higher idle power consumption of approximately 3.4~W. However, the power consumption remains relatively stable as the bandwidth increases. By optimizing the polling parameters (Figure~\ref{fig:power}(b)(d)), such as increasing the batch size to 40 or the packet size to 1500 bytes, KeepON's idle power is reduced to 2.7~W, and KeepON becomes more energy-efficient than GENET at target bandwidths exceeding approximately 300\,Mbps. This demonstrates that while KeepON's polling-based mechanism inherently consumes more power in an idle state, it becomes more efficient under high network loads by avoiding the significant overhead of frequent interrupts. Furthermore, this advantage is maximized through careful tuning of its PMD parameters.

\begin{figure}[tb]
    \centering
    \vspace{-0.1in}
    \includegraphics[width=.45\textwidth]{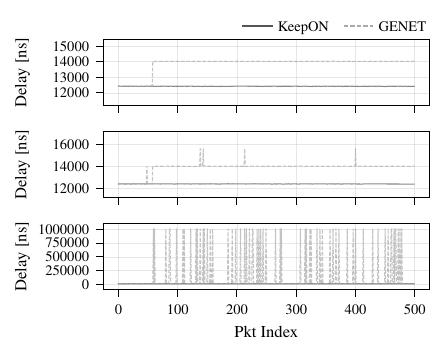}
    \caption{Time series analysis showing cascading jitter propagation in GENET vs. deterministic timing in KeepON.}
    \label{fig:series}
    \vspace{-0.15in}
\end{figure}

\vspace{0.02in}
\noindent\textbf{Case Study Analysis.} 
To investigate the cause of out-of-bound delay and jitter in GENET, Figure~\ref{fig:series} plots the schedule-to-delivery delay of the first 500 packets. The results reveal a domino effect in GENET’s jitter propagation, where a single misplaced packet cascades to subsequent scheduling. For example, when F2 misses its scheduled slot and occupies F3’s slot, F3 displaces F4, forcing F4 to wait until the next hyperperiod, resulting in a 1-ms delay spike. This disruption persists across hyperperiods: F2 continues to use F3’s slot because F4 from the previous hyperperiod occupies F2’s original slot, creating a permanent +2\,$\mu$s delay for F2. This behavior occurs because TSN’s queuing-based scheduling relies on temporal isolation from previous hops; when multiple flows share the same queue, timing violations at the end system propagate downstream, making the network vulnerable to such disruptions. 

\subsection{CPT Flow Specifications}
\label{sec:appendix-flows}

Table~\ref{tab:testbed-flow-param} details the 12 critical flows evaluated in our TSN testbed experiments. Flows marked with asterisks (*) represent legacy devices that lack native TSN support and require gateway integration through Raspberry Pi 4B devices running either KeepON or GENET drivers. These gateway-mediated flows (F2-F4, F7-F8) are particularly important for evaluating KeepON's performance, as they demonstrate the challenge of integrating non-TSN devices into a deterministic network. The remaining flows originate from TSN-capable end-stations with hardware scheduling offloading, providing a baseline for comparison.

Our end-to-end measurements are limited to flows F1-F4 and F7-F8 F12 that terminate at the RT simulator, as it is equipped with IEEE 1588 hardware timestamping capability for accurate MAC-layer delay measurement. Flows F5-F6 and F9-F10, which terminate at Raspberry Pi gateways, cannot be accurately measured due to the lack of hardware timestamping support on the BCM2711 NIC, which would introduce significant measurement uncertainty. 

\input{figs/testbed/flows.tex}

\subsection{System-Level Optimization}
We use the line rate of 1\,Gbps for all experiments, which is the maximum for RPi's default NIC. We use the default energy governor, such as CPU frequency scaling and idle state power adaptation. We set the energy mode as on-demand for Figure~\ref{fig:power} to observe true power consumption. 
We set the Linux socket buffer to a large value of 250MB to avoid potential packet loss. To reduce the impact of the system scheduler and accurately show how PMD parameters affect system performance, we isolate one or two cores to run the polling thread of KeepON for single-clock and dual-clock mode, respectively, affixing the PMD thread to each core. 
We disable the CPU periodic timer interrupt and IRQ for isolated cores. 
The single clock mode for KeepON is applied unless otherwise specified. 
We use default driver settings for GENET and IGB, and same system settings for all unless specified otherwise, consistently in our evaluations. By default setting, in the baseline GENET driver, TX interrupts are coalesced with a default timeout of 5\,ms. The IGB driver uses adaptive interrupt moderation which dynamically adjusts the interrupt rate based on traffic patterns.

\subsection{Traffic Isolation: Model and Constraints}
\label{ssec:isolation_model}

\noindent \textbf{Traffic scheduling model.}
\vspace{0.01in}
\noindent \textit{1) DMA Ring Buffer:} A ring buffer $B$ is equally divided into $N$ descriptor slots $s\in \{0, 1, \dots, N-1\}$. %
The transmission time of each slot is constant as $\delta$ (depending on the slot size), and slot $s$ is available for a transmission to start as $t$, when $s \equiv \lfloor t/\delta \rfloor\mod N$. We assume each slot only handles one packet transmission.

\vspace{0.01in}
\noindent \textit{2) Critical Apps:} A critical application set $\mathcal{A} = \{A_1, A_2, \dots, A_K\}$ share the single buffer with best-effort applications. 
Each critical application $A_i$ runs a set of real-time flows. Each flow $j\in A_i$ is characterized by its packet release period $p_{i,j}$ and maximum allowed jitter $j_{i,j}$ relative to the release period. Assume $p_{i,j}$ is multiples of $\delta$.
The hyperperiod $H$ is the least common multiple of all the flow periods in all the critical applications, {\em i.e.}, $H = LCM(\{p_{i,j} \mid A_i \in \mathcal{A}, j \in A_i\})$.

\vspace{0.01in}
\noindent \textit{3) Partitions:} A partition set $\mathcal{F} = \{f_1, f_2, \dots, f_K\}$ consists of disjoint partitions each of which corresponds to a subset of slots in the ring buffer, {\em i.e.}, $f_i \subseteq B$, and is allocated to $A_i$.

\vspace{0.02in}
\noindent \textbf{Per-application schedule definition.} A schedule $\mathcal{S}_i$ is defined for each application $A_i$ specifying the transmission time for all its flows in the hyperperiod using the slots in partition $f_i$. 
Each flow $j$ releases $M_j = H / p_{i,j}$ instances in the hyperperiod, and the transmission time for any instance, denoted as $t_{i,j,l}$, is determined by schedule $\mathcal{S}_i$. Correspondingly, the slot used for transmitting the instance is $s_{i,j,l} = \lfloor t_{i,j,l}/\delta\rfloor \bmod{N}$.

\vspace{0.02in}
\noindent \textbf{Valid schedule.} A schedule $\mathcal{S}_i$ generated for application $A_i$ using the allocated partition $f_i$ is valid if $\mathcal{S}_i$ satisfies the following constraints for all flows $j \in A_i$:

\noindent
1) The slot used for transmitting each instance of application $A_i$'s flows must be within the partition allocated to $A_i$, {\em i.e.}, 
\begin{equation}
    \forall i,j,l, s_{i,j,l} = \lfloor t_{i,j,l}/ \delta\rfloor \bmod{N} \in f_i.
\label{eq:cons1}
\end{equation}

\noindent 2) The time difference between any two packet transmissions ($|t_{i,j,a} - t_{i,j,b}|$) must not deviate from the difference in their specified release times ($|(a - b)|\cdot p_{i,j}$) by more than the jitter $j_{i,j}$, {\em i.e.}, %
\begin{equation}
    \forall i,j,a\neq b, \left | |(t_{i,j,a} - t_{i,j,b})| - |(a - b)|\cdot p_{i,j} \right |\leq j_{i,j}.
\label{eq:cons2}
\end{equation}

\vspace{0.01in}
\noindent 3) Any two packets within the same application cannot transmit at the same time, {\em i.e.}, $\forall i,j\neq j',l\neq l', t_{i,j,l}\neq t_{i,j'l'}$.

The partitioning problem is to find a feasible partition set $\mathcal{F}$ such that at least one valid schedule $\mathcal{S}_i$ exists for each $A_i$ using partition $f_i$. 

\begin{figure}[tb]
    \centering
    \includegraphics[width=0.45\textwidth]{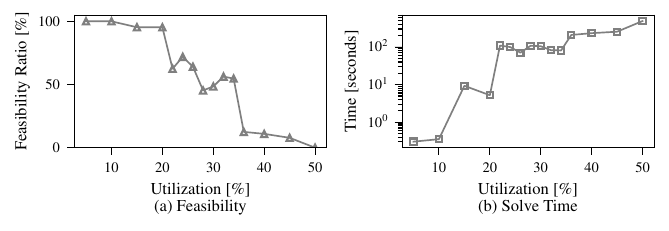}
    \caption{Schedulability analysis across utilization levels. (a) Percentage of flowsets with feasible isolation and schedules. (b) Average Z3 solving time showing exponential growth with utilization.}
    \label{fig:feasibility}
\end{figure}

\subsection{Isolation Configuration}
\label{ssec:isolation_appendix}

To evaluate the feasibility and tractability of our traffic isolation scheduling model, we conducted extensive simulation experiments using the Z3 SMT solver to analyze schedulability under varying real-time traffic loads.

\vspace{0.02in}
\noindent\textbf{Experimental Setup.}
We generate synthetic flowsets with system utilization ranging from 5\% to 50\%. The utilization is calculated as the ratio of total packet instances per ring cycle to the number of slots. Given each utilization value, we created 64 different flow configurations using the parameters summarized in Table~\ref{tab:flowset-params}.

\begin{table}[htbp]
\small
\centering
\caption{{Flowset parameters for schedulability analysis}}
\label{tab:flowset-params}
\begin{tabular*}{0.48\textwidth}{@{} l @{\extracolsep{\fill}} l @{}}
\toprule
\textbf{Parameter} & \textbf{Configuration} \\
\midrule
System utilization & \{5\%, 10\%, 15\%, ..., 50\%\} \\
Number of flows & \{2, 4, 8, 12, 16\} \\
Flow periods & \{80, 160, 240, ..., 1600\,$\mu$s\} \\
Max jitter & 5 -- 20\% of flow period \\
Slot size & 10\,$\mu$s \\
Ring size & 32 slots \\
Traffic classes & $\leq$ 8 classes \\
Datasets per experiment & 64 \\
\bottomrule
\end{tabular*}
\end{table}

\vspace{0.02in}
\noindent\textbf{Z3 Scheduling Analysis.}
For each flowset, we employed Z3 to determine whether a feasible partition $\mathcal{F}$ and schedule $\mathcal{S}_i$ exist that satisfy all constraints (in Section~\ref{ssec:isolation}). The solver searches for slot allocations to traffic classes and transmission times within the hyperperiod, with a 150-second timeout for complex instances. If the solver cannot find a feasible solution within the given time period limit, the flowset is considered infeasible. All flow periods are chosen as harmonic to ensure bounded hyperperiods. 

\vspace{0.02in}
\noindent\textbf{Results.}
Figure~\ref{fig:feasibility} presents the schedulability results across 896 flowsets:%

\vspace{0.01in}
\noindent \textit{1) Schedulability:} As shown in figure~\ref{fig:feasibility}(a), the schedulability reaches almost 100\% at low utilization (5-20\%) and decreases sharply when the utilization is larger than 25\%. %
This demonstrates that our isolation model effectively handles typical industrial real-time workloads while revealing the fundamental limits of static partitioning.

\vspace{0.01in}
\noindent \textit{2) Runtime:} As shown in figure~\ref{fig:feasibility}(b), Z3 solving time increases exponentially (less than 0.5 seconds at 5\% utilization and over 100 seconds at 40-50\% utilization), reflecting the growing constraint complexity as the solution space becomes more constrained.

\subsection{PTP Performance Analysis}
\label{ssec:ptp-analys}
To further understand the synchronization performance, we analyze the maximum time offset between the master and slave clocks caused by inaccurate timestamping at a single hop.
We consider two sources of timestamping uncertainty: 1) Random timestamping delay, i.e., $j_M^{in}$, $j_M^{out}$, $j_S^{in}$, and $j_S^{out}$; 2) Clock granularity, denoted as $g_M$, and $g_S$, representing the resolution of the master and slave clocks, respectively. 
Below, we analyze the maximum offset inaccuracy ($\delta_{TS}$) and maximum drift accumulation ($\delta_{Drift}$) caused by potentially inaccurate timestamping. 

\vspace{0.02in}
\noindent $\bullet$ \textit{Offset Inaccuracy:} The PTP slave corrects its clock based on the offset between the master and slave clocks. The offset is calculated as $O = [(t_2 - t_1) - (t_4 - t_3)] / 2$, as shown in Fig.~\ref{fig:ptp} (\ding{192}). 
Ideally, the corrected time matches the master's time without any offset. In practice, inaccurate timestamping causes each timestamp to deviate from the true event time $t_{i, est} = t_{i, true} + \epsilon_i$. The error in this offset calculation is $\left((\epsilon_{2}-\epsilon_{1}\right)-\left(\epsilon_{4}-\epsilon_{3}\right))/2$.

The error range for each timestamp is based on its granularity and jitter. That is:
\begin{equation}
\begin{aligned}
    & \epsilon_{1} \in\left[-\frac{1}{2} g_M, \frac{1}{2} g_M+j_M^{{out }}\right] \quad && \epsilon_{2} \in\left[-\frac{1}{2} g_S, \frac{1}{2} g_S+j_S^{{in }}\right] \\
    & \epsilon_{3} \in\left[-\frac{1}{2} g_S, \frac{1}{2} g_S+j_S^{{out }}\right] \quad && \epsilon_{4} \in\left[-\frac{1}{2} g_M, \frac{1}{2} g_M+j_M^{{in }}\right]
\end{aligned}
\label{eq:timestamp_error}
\end{equation}

The maximum absolute value of the offset calculation error, which represents the uncertainty in the offset correction applied by the slave, can then be bounded by: 
\begin{equation}
\delta_{T S}=\frac{1}{2}\left(g_M+g_S+\max \left(j_M^{\text {in }}+j_M^{\text {out }}, j_S^{\text {in }}+j_S^{\text {out }}\right)\right)
\label{eq:max_sync_offset}
\end{equation}

\vspace{0.02in}
\noindent $\bullet$ \textit{Drift Accumulation:} Even with frequency correction (\ding{193}), a residual frequency error persists due to timestamp errors in frequency ratio estimation. 
The frequency ratio is estimated as $(t_1^{next} - t_1)/(t_2^{next} - t_2)$, and the measured ratio casued by timestamp errors is:
\begin{equation}
r_{meas} = \frac{\Delta t_M + \epsilon_M}{\Delta t_S + \epsilon_S}
\end{equation}
where $\epsilon_M \in [-(g_M + j_M^{out}), g_M + j_M^{out}]$ and $\epsilon_S \in [-(g_S + j_S^{in}), g_S + j_S^{in}]$ represent the timestamp error differences between consecutive sync messages.

After frequency correction using $r_{meas}$, the residual frequency difference is $\Delta\rho = \rho_S(r - r_{meas})$, where $r = \rho_M/\rho_S$ is the true ratio. 
The worst-case drift occurs when $r_{meas}$ deviates maximally from $r$. 
For synchronized clocks where $\Delta t_M \approx \Delta t_S \approx I$, the maximum drift over interval $I$ is:
\begin{equation}
\begin{aligned}
& \delta_{Drift} = I \times \max(|\Delta \rho_{max}|, |\Delta \rho_{min}|) \\
&= I \times \rho_M \times \max\left[\frac{g_S + j_S^{in}}{I - g_S - j_S^{in}}, \frac{g_M + j_M^{out}}{I + g_S + j_S^{in}}\right]
\end{aligned}
\end{equation}
The first term represents drift from overestimating the frequency ratio when master timestamps have positive errors and slave timestamps have negative errors, 
while the second term represents underestimation. 

\begin{figure}[tb]
    \centering
    \includegraphics[width=0.45\textwidth]{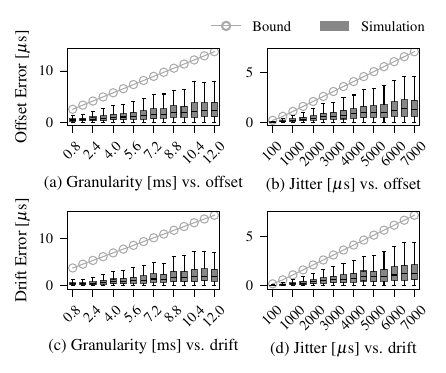}
    \caption{{Impact of timestamp error parameters on PTP synchronization accuracy in simulation.}}
    \label{fig:simu_ptp_impact}
\end{figure}

\begin{figure}[tb]
    \centering
    \includegraphics[width=0.45\textwidth]{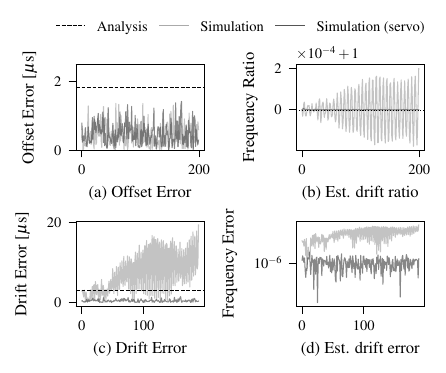}
    \caption{{Comparison of PTP synchronization in simulation with and without frequency filtering.}}
    \label{fig:simu_ptp_error}
\end{figure}

\vspace{0.02in}
\noindent \textbf{Simulation Results.}
To validate our theoretical analysis, we implement a PTP synchronization simulator with the following settings: a synchronization interval of 100\,ms, a symmetric network delay of 10\,µs, and a 1\,ppm frequency offset between the master and slave clocks in initialization. We use a moving average window size of 10 to smooth the frequency in the simulation. 
Figure~\ref{fig:simu_ptp_impact} shows the impact of timestamping jitter and clock granularity (decided by slot-size) on synchronization performance. 
As shown in Figure~\ref{fig:simu_ptp_impact}(a,c), clock granularity increases from 0.8 to 12\,$\mu$s, both offset and drift errors grow linearly, with median offset errors increasing from 0.5 to 2.7\,$\mu$s as 5.4$\times$ increase. 
As shown in Figure~\ref{fig:simu_ptp_impact}(b,d), jitter has a more pronounced effect as its offset error increases from 0.03 to 1.5\,$\mu$s by 50$\times$, demonstrating that jitter has more impact on synchronization accuracy. 

It is worth noting that our analysis result bounds the maximum drift within a single synchronization interval but does not account for this multi-interval accumulation effect. Figure~\ref{fig:simu_ptp_error}(b,c,d) grey line simulates a common limitation in PTP's drift correction that the estimated frequency ratio oscillates randomly due to random timestamping delay, which eventually causes drift errors to grow unbounded. 
To address this accumulation, many real PTP implementations apply different servo algorithms to filter the frequency ratio estimate.
In our simulation, we implemented a simple 10-sample moving average filter to smooth the frequency.
With filtering (black line in Figure~\ref{fig:simu_ptp_error}(b,c,d)), the frequency ratio converges smoothly to the true value, reducing average drift errors by 12.2$\times$ and keeping them near the single-interval bound. 
This demonstrates that while the theoretical analysis accurately predicts per-interval errors, practical implementations require frequency filtering to prevent the random walk accumulation that occurs over multiple synchronization cycles.

\subsection{Parameter Selection}
The selection of CP-PMD parameters (\textit{i.e.}, slot size and batch size) has great impact on both real-time performance (feasibility, jitter, latency, clock accuracy) and system performance (CPU usage, bandwidth, energy efficiency). This section provides a practical guideline to assist users in selecting appropriate parameters, with the objective of satisfying real-time constraints while minimizing system overhead. 
Our discussion uses the integration with a TSN network as an example.

\vspace{0.1in}
\noindent\textbf{Trade-off Analysis.}  
Table~\ref{tab:param-impact} summarizes how the two primary CP-PMD parameters (\textit{i.e.},
slot-size and batch-size) affect key system metrics, which are elaborated below.

\vspace{0.02in}
\bulletitem{\textit{Time Granularity}} (\S\ref{ssec:EPHC}): The time granularity
equals the slot transmission time $\delta = \text{slot-size} /
\text{line-rate}$, so a larger slot-size directly coarsens the
time granularity. Batch-size does not affect $\delta$.

\vspace{0.02in}
\bulletitem{\textit{Advance Time}} (\S\ref{sec:scheduled-packet-insertion}): A data packet must
arrive at the driver before the start of the immediate insertion window,
requiring a minimum lead time of $L_{\min} = \text{batch-size} \times
\delta$. Since $\delta$ is proportional to slot-size, both parameters
linearly increase the advance time and thus the data age at transmission.

\vspace{0.02in}
\bulletitem{\textit{Synchronization Accuracy}} (\S\ref{sec:sync}, Eq.~\ref{eq:max_sync_offset}):
The PTP offset bound $\delta_{TS}$ depends on clock granularity $g$, which grows linearly with slot-size. For batch-size, the EPHC is
updated only once per polling iteration, so the effective granularity
becomes $g_{\text{eff}} = \text{batch-size} \times \delta$; however, the
degradation is masked by other system jitter until $g_{\text{eff}}$
exceeds a threshold, producing the abrupt transition observed in
Figure~\ref{fig:offset}(b).

\vspace{0.02in}
\bulletitem{\textit{Bandwidth Efficiency}} (\S\ref{ssec:bw}): Fixed-size slots 
require padding when the payload $P$ is smaller than the slot-size, yielding a
payload efficiency of $P / \text{slot-size}$. Larger slot-sizes
therefore waste more bandwidth on padding. Batch-size does not affect the
data-to-padding ratio on the wire.

\vspace{0.02in}
\bulletitem{\textit{CPU Efficiency}} (\S\ref{ssec:cpu}): The polling budget $\text{batch-size} \times \delta$ determines the time
available between consecutive PMD iterations~(\S\ref{ssec:driver-impl}). When budget is
smaller than the per-batch processing overhead, the polling thread
saturates the CPU core; once budget exceeds this overhead, idle
time appears and utilization drops sharply, exhibiting threshold behavior
for both parameters (Figure~\ref{fig:cpu}).

\vspace{0.02in}
\bulletitem{\textit{Power Efficiency}} (\S\ref{ssec:energy}): Since KeepON's
power consumption is primarily determined by CPU utilization of the
polling thread, power efficiency follows the same relationship as CPU
utilization for both parameters. Experiments in
Figure~\ref{fig:power} confirms this relationship.

\begin{table}[t]
\centering
\caption{Impact of slot-size and batch-size on system metrics. $\uparrow$/$\downarrow$: metric increases/decreases with larger parameter value; $-$: no direct effect; (thres): exhibits threshold behavior.}
\label{tab:param-impact}
\begin{tabular*}{\linewidth}{@{\extracolsep{\fill}}lll@{}}
\toprule
\textbf{Metric}          & \textbf{Slot-size} & \textbf{Batch-size} \\ \midrule
Time Granularity       & $\uparrow$              & $-$                     \\
Advance Time (Delay)     & $\uparrow$              & $\uparrow$              \\
Synchronization Accuracy & $\downarrow$            & $\downarrow$ (thres)    \\
Bandwidth Efficiency     & $\downarrow$            & $-$                     \\
CPU Efficiency          & $\uparrow$ (thres)    & $\uparrow$ (thres)    \\
Power Efficiency         & $\uparrow$              & $\uparrow$              \\ \bottomrule
\end{tabular*}
\end{table}

\vspace{0.1in}
\noindent\textbf{Parameter Selection.} 
Slot size and batch size jointly determine KeepON's timing accuracy and system overhead (Table~\ref{tab:param-impact}).
We recommend selecting them sequentially: first the slot size (as it affects the isolation model), and then the batch size.

\textit{Slot-size} should be as small as possible to preserve finer time granularity, synchronization accuracy and bandwidth, 
but large enough to avoid fragmenting packets. 
A good starting point is to pick candidates that are multiples of the GCD of all real-time flow periods, 
which minimizes the hyperperiod and improves scheduling feasibility (\S\ref{ssec:isolation_appendix}).
Among these candidates, choose the smallest one that is at least as large as the maximum packet size, so no fragmentation is needed. Then verify that the resulting synchronization error (Eq.~\ref{eq:max_sync_offset}) stays within the network's tolerance. If the isolation feasibility test (\S\ref{ssec:isolation_appendix}) fails at this slot-size, reduce it further (accepting that some packets will need fragmentation) and re-check the sync error bound.

Selecting \textit{batch-size} is simpler since it does not affect the isolation model. We recommend choosing the largest batch-size such that: a) the synchronization error (Eq.~\ref{eq:max_sync_offset}) remains within budget, and b) batch-size does not exceed the application's maximum tolerable data age. Within that range, the user can increase batch-size until the CPU utilization decreases to an acceptable value.

\subsection{Additional Discussion}

\noindent\textbf{Scheduling Accuracy.}
A counter-intuitive finding in \S\ref{sec:evaluation} is that KeepON's software pacing achieves tighter timing precision (3.7\,ns) than the Intel i210's hardware offload (7.7\,ns).
This gap stems from how the i210's LaunchTime mechanism makes its release decision: the NIC uses a digital time comparator that matches the current system time against the programmed launch timestamp at a discrete resolution.
Specifically, the i210 compares \texttt{SYSTIML[29:5]} (i.e., it ignores the lower 5 bits) yielding a transmission-time granularity of 0.032\,$\mu$s (Sec.~7.2.7.5.3~\cite{intel-i210-datasheet}).
In contrast, KeepON does not depend on a discrete timer comparison.
Instead, it aligns packet transmissions to the continuous physical flow of bytes on the wire by keeping the link saturated using CP-PMD and inserting real packets by overwriting placeholders.
At a 1\,Gbps line rate, transmitting one byte takes 8\,ns, so the pacing mechanism naturally operates at an 8\,ns byte-time granularity, enabling finer timing alignment than the i210's comparator-based LaunchTime offload.

\vspace{0.02in}
\noindent\textbf{Isolation by DMA Partition.} Conventional drivers isolate traffic classes using hardware multi-queueing (e.g., \texttt{mqprio}), mapping traffic with different levels of real-time requirements (e.g., best-effort, AVB, critical traffic) to separate TX queues. In contrast, CP-PMD achieves accurate scheduling by continuously polling a single DMA ring to sustain strict line-rate pacing, making multi-queue isolation incompatible with KeepON's driver model. Instead, we exploit KeepON's temporal and spatial relationship: as discussed in \S~\ref{ssec:EPHC}, a packet's ring index determines when it is transmitted. Thus, partitioning the ring into disjoint index ranges provides isolation by partitioning the corresponding transmission time windows.

\vspace{0.02in}
\noindent\textbf{Portability.} We implement the prototype of KeepON based on Raspberry Pi's bcmgenet driver. The implementation added about 1100 lines of code, primarily in four areas: 1) implementing TX data path and scheduling logic ($\sim$200 lines), 2) adding EPHC and dual-clock logic ($\sim$500 lines), 3) CP-PMD polling and initialization ($\sim$200 lines), and 4) modifying driver interfaces ($\sim$200 lines). The underlying code about hardware initialization and PHY layer interactions remain largely untouched. Since most modern NIC drivers share a similar ring-descriptor architecture, the core logic of KeepON is driver-agnostic and broadly applicable to other drivers.

\vspace{0.02in}
\noindent\textbf{Applicability.} KeepON is best suited for scenarios where two conditions are met. First, deploying a specialized NIC is impractical, either because the device lacks a PCIe expansion slot (e.g., Raspberry Pi), or the platform is already certified and cannot be modified, or the per-NIC cost is prohibitive at scale. Second, the device has idle CPU cores available for the PMD polling thread. This is typical of network gateways where the primary workload is communication rather than computation, such as the RPi gateways in our case study (\S\ref{sec:case-study}) that bridge legacy serial and analog devices to the TSN backbone. For other cases, we recommend hardware offloading approaches using specialized NICs~\cite{coleman2019emerging, xue2024towards}.

\subsection{API Design}
We implement the Synchronization module and the Traffic Management module with existing Linux APIs to provide easy-to-use interfaces for seamless integration.

\vspace{0.02in}
\noindent \textbf{Timestamp delivering.} 
To keep compatibility with existing Linux APIs, we use the standard \texttt{SO\_TXTIME} socket option to deliver the scheduled transmission time from user-space, as shown in Listing~\ref{lst:setsockopt}. The socket option conveys the desired timestamp via control messages \texttt{CMSG} within the \texttt{sendmsg} call, and the driver can parse the timestamp from the kernel's socket buffer (skb) structure. 

\begin{figure}[htbp]
\begin{lstlisting}[style=mystyle, language=C, caption={API: Example of userspace application to enable \texttt{SO\_TXTIME} socket option and send scheduled packet.}, label={lst:setsockopt}]
// Enable SO_TXTIME socket option
setsockopt(fd, SOL_SOCKET, SO_TXTIME, &sk_txtime,
sizeof(sk_txtime));

// Set scheduled time in CMSG
cmsg = CMSG_FIRSTHDR(&msg);
cmsg->cmsg_level = SOL_SOCKET;
cmsg->cmsg_type = SO_TXTIME;
cmsg->cmsg_len = CMSG_LEN(sizeof(__u64));
*((__u64 *) CMSG_DATA(cmsg)) = txtime;

// Send message
ret = sendmsg(fd, &msg, 0);
\end{lstlisting}
\end{figure}

\vspace{0.02in}
\noindent \textbf{Pre-Buffering.} 
We utilize the queuing discipline (qdisc) framework in the Linux Traffic Control (TC) subsystem to implement traffic pre-buffering and prioritization~\cite{tc-etf}. 
First, we use the \texttt{mqprio} qdisc to create multiple logical traffic classes depending on the number of applications. While \texttt{mqprio} maps traffic classes to multiple hardware queues, in our configuration, these logical classes are mapped to partitions (\S\ref{ssec:isolation}) of a single hardware queue. 
Listing~\ref{lst:prebuffer} shows an example: \ul{\textit{1) For real-time traffic:}} (e.g., \texttt{1:3} and \texttt{1:4}), we attach an Earliest Time First (\texttt{etf}) qdisc to each respective class. This qdisc buffers packets and dequeues them based on their \texttt{SO\_TXTIME} timestamp, which is set by the application to indicate the desired transmission time. 
The \texttt{etf} qdisc uses an efficient data structure, often a red-black tree, for ordering packets by their release time. The \texttt{delta} parameter specifies an offset before the packet's scheduled timestamp when it should be released to the network driver. 
This accounts for worst-case packet transmisison latencies, e.g., caused by system jitter and batching effects.
\ul{\textit{2) For best-effort traffic:}} (e.g., traffic mapped to parents such as 1:1 and 1:2 in a four-class setup), a standard \texttt{pfifo} qdisc is attached to each of these classes. This qdisc provides basic queuing without prioritization beyond the initial class assignment, ensuring that packets in the order they are received within their class.
Traffic classification is determined at the application layer: real-time packets are identified by their \texttt{SO\_TXTIME} timestamp set via the socket interface, while packets without a designated timestamp are treated as best-effort and routed through standard Linux traffic control (qdisc) class mapping.

\begin{figure}[htbp]
\begin{lstlisting}[style=mystyle, language=bash, caption={API: Example for pre-buffering set up by Linux Qdisc.}, label={lst:prebuffer}]
## Application 0-7 use class 0, 8 use class 1
## 9-11 use class 2, 12-15 use class 3
sudo tc qdisc add dev eth0 root handle 1: mqprio num_tc 4 \
    map 0 0 0 0 0 0 0 1 2 2 2 2 3 3 3 3 \
    queues 1@0 1@1 1@2 1@3 \
    hw 0

## Set class 1-2 as FIFO qdisc
sudo tc qdisc replace dev eth0 parent 1:1 pfifo
sudo tc qdisc replace dev eth0 parent 1:2 pfifo

## Set class 3-4 as ETF qdisc
sudo tc qdisc replace dev eth0 parent 1:3 etf clockid CLOCK_TAI delta 50000
sudo tc qdisc replace dev eth0 parent 1:4 etf clockid CLOCK_TAI delta 50000
\end{lstlisting}
\vspace{-0.2in}
\end{figure}

\vspace{0.02in}
\noindent \textbf{Synchronization.} To integrate with standard Linux time synchronization mechanisms, our driver exposes its EPHC capabilities through the PTP Hardware Clock (PHC) kernel API. 
This creates character devices in the \texttt{/dev/} directory (e.g., \texttt{/dev/ptp0}, \texttt{/dev/ptp1}), allowing userspace tools to access and control these hardware clocks. The specific \texttt{/dev/ptpX} device corresponding to the network interface's timestamps (e.g., a TX-EPHC in single-clock setup or a combined-EPHC in dual-clock setup) is used by the PTP demon for internal or external synchronization. Listing~\ref{lst:ptp} shows a typical synchronization setup with default settings.

\vspace{0.02in}
\noindent$\bullet$ \textit{phc2sys} synchronizes the system clock (e.g., \texttt{CLOCK\_REALTIME}) to the internal PTP hardware clock (\texttt{/dev/ptpX}).

\vspace{0.02in}
\noindent$\bullet$ \textit{ptp4l} implements the IEEE 1588 PTP protocol to synchronize the hardware clock (\texttt{/dev/ptpX}) with an external PTP clock over the network.

\begin{figure}[htbp]
\begin{lstlisting}[style=mystyle, language=bash, caption={API: Example of synchronization setup.}, label={lst:ptp}]
# Master
sudo phc2sys -s /dev/ptp0 -O 0 -m
sudo ptp4l -i eth0 -m -P

# Slave
sudo phc2sys -s /dev/ptp0 -O 0 -m
sudo ptp4l -i eth0 -m -P -s -p /dev/ptp1
\end{lstlisting}

\end{figure}

\vspace{0.02in}
\noindent \textbf{Driver Parameters.} 
KeepON driver's PMD behaviors can be configured through module parameters during initialization. Key parameters include:

\vspace{0.02in}
\noindent$\bullet$ \textit{slot\_masks}: Specifies slot allocation for each traffic class using bitmasks. Each bit represents a slot in the fixed-size (by default 32) ring buffer. For example, Listing~\ref{lst:module_params}  allocates slot 0 to class 0 and slots 1,17 to class 1, with the pattern repeating every 3.2ms cycle (32 slots $\times$ 100$\mu$s).

\vspace{0.02in}
\noindent$\bullet$ \textit{pkt\_size}: Sets the slot size in bytes (up to 1518). This determines the line-rate packet transmission time.

\vspace{0.02in}
\noindent$\bullet$ \textit{batch\_size}: Controls how many descriptors the polling thread processes per iteration (1-512), affecting CPU efficiency and synchronization performance.

\vspace{0.1in}
\begin{lstlisting}[style=mystyle, language=bash, caption={API: Example of loading KeepON driver.}, label={lst:module_params}]
# Load driver with two real-time classes
sudo insmod genet.ko slot_masks=0x01,0x20002,0,0,0 pkt_size=1230 batch_size=8
\end{lstlisting}

%% file: figs/testbed/flows.tex
\begin{table}[]
    \caption{Critical flows in the experimental testbed. Asterisks (*) indicate legacy devices connected through RPi gateway rather than direct TSN-capable end-stations.}
    \label{tab:testbed-flow-param}
    \resizebox{\columnwidth}{!}{
    \begin{tabular}{@{}lllll@{}}
    \toprule
    \textbf{FlowId} & \textbf{Flow Path} & \textbf{Payload Description} \\ \midrule
    1  & Power Load → RT simulator        & Load Telemetry (V/I/P, status)           \\
    2  & Power Supply* → RT simulator     & Supply Telemetry (V/I, status)           \\
    3  & Power Amplifier* → RT simulator  & Amplifier Telemetry (gain, temp)         \\
    4  & PV simulator* → RT simulator     & PV simulator telemetry (V/I curve)       \\
    5  & RT simulator → Power Load        & Setpoints (current/voltage, enable)      \\
    6  & RT simulator → Power Amplifier*  & Commands (amplitude/gain, enable)        \\
    7  & Pressure Sensor* → RT simulator  & Pressure readings                        \\
    8  & Thermocouple* → RT simulator     & Temperature readings                     \\
    9  & RT simulator → Heatpad*          & Setpoints (power/temperature)            \\
    10 & RT simulator → Compressor*       & Commands (speed/duration)                \\
    11 & Structural Sensor → GPU workstation & Dense spatial vibration data          \\
    12 & GPU workstation → Structural Sensor & State estimates                       \\ 
    \bottomrule
    \end{tabular}}
\end{table}

%% file: 10-ae-appendix.tex
\section*{Artifact Appendix}

This artifact provides a prototype implementation of KeepON on the Broadcom bcmgenet NIC driver of the Raspberry Pi 4B. The artifact includes the modified kernel driver source code with kernel 6.14.7 and user space test examples that enable deterministic packet transmission on standard NICs. The artifact is publicly available on Github: \url{https://github.com/ChuanyuXue/KeepON-rpi}

\subsection{Build}

Compile as module:

\begin{lstlisting}[style=mystyle, language=bash]
make -C /lib/modules/$(uname -r)/build M=$(pwd) \
  modules EXTRA_CFLAGS="-DENABLE_DUAL_CLOCK_MODE=0"
\end{lstlisting}

\noindent \textbf{Note:} \texttt{/drivers/net/ethernet/broadcom/unimac.h} must be copied to an upper-level path when compiling as a module.

\begin{lstlisting}[style=mystyle, language=bash]
ethtool -T eth0
# Time stamping parameters for eth0:
# Capabilities:
#   hardware-transmit
#   software-transmit
#   hardware-receive
#   software-receive
#   software-system-clock
#   hardware-raw-clock
# PTP Hardware Clock: 0
\end{lstlisting}

\vspace{0.02in}
\noindent After successful installation, you should see the emulated PTP hardware clock (EPHC):

\subsection{Examples}

\vspace{0.02in}
\noindent\textbf{Example 1: Real-Time Traffic Only}

\begin{lstlisting}[style=mystyle, language=bash]
# Allocate all slots to tc-0
sudo insmod genet.ko \
  slot_masks=0xffffffff,0,0,0,0 pkt_size=1230

# Configure pre-buffering heap for tc-0
sudo tc qdisc add dev eth0 root handle 1: \
  mqprio num_tc 2 \
  map 1 1 1 1 1 1 1 1 0 0 0 0 0 0 0 0 \
  queues 1@0 1@1 hw 0

sudo tc qdisc replace dev eth0 parent 1:1 \
  etf clockid CLOCK_TAI delta 150000

# Synchronize system clock
sudo systemctl stop systemd-timesyncd
sudo phc2sys -s /dev/ptp0 -O 0 -m

# Send strictly periodic traffic
sudo ./test/cyclic 100000 200000 100
\end{lstlisting}

\noindent Receiver should see strictly 100$\mu$s period:

\begin{lstlisting}[style=mystyle, language=bash]
sudo tcpdump -Q in -ttt -ni eth0 \
  --time-stamp-precision=nano -j adapter_unsynced

# 00:00:00.000100000  [|ether]
# 00:00:00.000100000  [|ether]
# 00:00:00.000099992  [|ether]
# 00:00:00.000100000  [|ether]
# 00:00:00.000100000  [|ether]
\end{lstlisting}

\vspace{0.02in}
\noindent\textbf{Example 2: Heterogeneous Traffic Management}

\begin{lstlisting}[style=mystyle, language=bash]
# Allocate for isolation
# tc-0: slot 0       (1 slot/cycle)
# tc-1: slot 1, 17   (2 slots/cycle)
# tc-2: remaining    (best-effort)
# Repeats every 3.2ms cycle (32 slots x 100us)

sudo insmod genet.ko \
  slot_masks=0x01,0x20002,0,0,0 pkt_size=1230

# Configure pre-buffering for tc-0, tc-1
sudo tc qdisc add dev eth0 root handle 1: \
  mqprio num_tc 3 \
  map 2 2 2 2 2 2 2 2 0 0 0 0 1 1 1 1 \
  queues 1@0 1@1 1@2 hw 0

sudo tc qdisc replace dev eth0 parent 1:1 \
  etf clockid CLOCK_TAI delta 150000
sudo tc qdisc replace dev eth0 parent 1:2 \
  etf clockid CLOCK_TAI delta 150000

# Synchronize system clock
sudo systemctl stop systemd-timesyncd
sudo phc2sys -s /dev/ptp0 -O 0 -m
\end{lstlisting}

\noindent tc-2 (best-effort): no guarantee on periodicity.

\begin{lstlisting}[style=mystyle, language=bash]
sudo ./test/cyclic 100000 200000 100 0
# 00:00:00.000140000  [|ether]
# 00:00:00.000099992  [|ether]
# 00:00:00.000120000  [|ether]
# 00:00:00.000080000  [|ether]
\end{lstlisting}

\noindent tc-0 (real-time): strictly 320$\mu$s period (1 slot per cycle).

\begin{lstlisting}[style=mystyle, language=bash]
sudo ./test/cyclic 100000 200000 100 8
# 00:00:00.000320000  [|ether]
# 00:00:00.000319992  [|ether]
# 00:00:00.000320000  [|ether]
# 00:00:00.000320000  [|ether]
\end{lstlisting}

\noindent tc-1 (real-time): strictly 160$\mu$s period (2 slots per cycle).

\begin{lstlisting}[style=mystyle, language=bash]
sudo ./test/cyclic 100000 200000 100 12
# 00:00:00.000160000  [|ether]
# 00:00:00.000160000  [|ether]
# 00:00:00.000159992  [|ether]
# 00:00:00.000160000  [|ether]
\end{lstlisting}

\vspace{0.02in}
\noindent\textbf{Example 3: TSN Testbed Integration}

\begin{lstlisting}[style=mystyle, language=bash]
# Talker (KeepON) <-> Switch (TTTech) <-> Listener (igb)
sudo insmod genet.ko pkt_size=1230

sudo tc qdisc add dev eth0 root handle 1: \
  mqprio num_tc 2 \
  map 1 1 1 1 1 1 1 1 0 0 0 0 0 0 0 0 \
  queues 1@0 1@1 hw 0

sudo tc qdisc replace dev eth0 parent 1:1 \
  etf clockid CLOCK_TAI delta 150000

# Synchronize with TSN grandmaster
sudo systemctl stop systemd-timesyncd
sudo phc2sys -s /dev/ptp0 -O 0 -m
sudo ptp4l -i eth0 -f gptp.cfg -m -P

# Send 1ms periodic traffic (PCP=5, VLAN=0)
sudo ./test/cyclic 1000000 200000 100 8 0 5 0
\end{lstlisting}

\vspace{0.02in}
\noindent Without TSN schedule, receiver observes $\sim$1\,ms period with $\mu$s-level variation. 

\begin{lstlisting}[style=mystyle, language=bash]
# 1520607476.439987235  [|ether]
# 1520607476.440987549  [|ether]
# 1520607476.441987511  [|ether]
# 1520607476.442987472  [|ether]
# 1520607476.443987426  [|ether]
# 1520607476.444987380  [|ether]
\end{lstlisting}

\noindent After enabling the TSN schedule on the TTTech switch:

\begin{lstlisting}[style=mystyle, language=bash]
# TSN schedule: 1ms cycle, 40us gate for queue 5
# test.cfg:
# sgs 40000 0x20  (queue 5, for real-time)
# sgs 40000 0x03  (queue 1-2, for BE and PTP)
tsntool st wrcl sw0p2 test.cfg
tsntool st configure 0.0 1/1000 0 sw0p2
\end{lstlisting}

\noindent Jitter reduces to $\leq$10\,ns:

\begin{lstlisting}[style=mystyle, language=bash]
# 1520607628.694000238  [|ether]
# 1520607628.695000240  [|ether]
# 1520607628.696000241  [|ether]
# 1520607628.697000243  [|ether]
# 1520607628.698000236  [|ether]
# 1520607628.699000238  [|ether]
\end{lstlisting}